\documentclass[a4paper,12pt]{article}
\usepackage{statcomp_archive}
\usepackage{hyperref}
\usepackage{amsfonts}
\usepackage{amssymb}
\usepackage{bm}
\DeclareMathAlphabet\mathbfcal{OMS}{cmsy}{b}{n}
\usepackage{rotating}

\title{Efficient Case-Cohort Design using Balanced Sampling}

\author[1]{Kaeum Choi}
\author[1,2]{Sangwook Kang\footnote{Corresponding author}}

\affil[1]{Department of Statistics and Data Science, Yonsei University}
\affil[2]{Department of Applied Statistics, Yonsei University}

\begin{document}
\maketitle

\clearpage
\begin{abstract}
 The case-cohort design is a cost-efficient two-phase design for analyzing survival data when key risk factors are expensive to assess and the event rate is low. Traditionally, subcohorts are selected via simple random sampling, which might not fully utilize available information. This paper introduces an efficient sampling design based on balanced sampling \citep{tille2006sampling} for a subcohort selection within the case-cohort design. A notable benefit of employing balanced sampling is the automatic calibration of auxiliary variables available for the entire cohort. Under a Cox model, it has been demonstrated that the calibration of sampling weights, utilizing auxiliary variables highly correlated with the main risk factor, significantly enhances the efficiency of regression coefficient estimators \citep{breslow2009improved,breslow2009using}. Extensive simulation experiments show the reduced variabilities under the proposed approach in comparison to those under both simple random sampling. The proposed design and estimation procedure are further illustrated using the well-established National Wilms Tumor Study dataset.

    
\end{abstract}

\clearpage
\pagenumbering{arabic}
\section{Introduction}\label{Sec:Introduction}

A case-cohort design \citep{prentice1986case} is a cost-efficient study design that conducts a sampling of subjects within the study cohort. 
It is a two-phase design that, in the first phase, the cohort is viewed as a random sample from a larger population, and, in the second phase, a subset called subcohort is chosen from the study cohort at the beginning of follow-up, augmented over time by all those who experience the event of interest, cases. Since expensive covariate information is collected exclusively for both the subjects in the subcohort and all the cases, the cost of conducting a large-scale cohort study can be reduced. Thus, this design is particularly useful in large cohort studies where it can efficiently investigate associations between exposures and outcomes without the need to analyze the entire cohort. Nevertheless, the aforementioned case-cohort design may have certain drawbacks when the subcohort is chosen through a simple random sampling. Randomly selecting a subcohort results in loss of important information in the cohort. A case-cohort sampling usually only includes a small portion of the non-cases, so it provides limited detailed information for the non-cases in the cohort. One way to address these limitations is utilizing auxiliary information by calibrating sampling weights \citep{deville1992calibration, sarndal2003model}. 


Calibration is the process of adjusting these weights to ensure that the total of the auxiliary variables that demonstrate a high correlation with variables of interest while being more cost-effective to measure and accessible for the entire cohort is exactly estimated. The rationale behind this process is that when the total of auxiliary variables is accurately estimated, it enhances the chance of accurately estimating the total of the variables of interest and recudes costs. There are many endeavors in applying this calibration approach in the context of enhancing the efficiency of case-cohort analysis \citep{breslow2009using,breslow2009improved, lumley2011connections}. In a two-phase design under a Cox model setting \citep{cox1972regression}, \citet{breslow2009using,breslow2009improved} proposed a calibration procedure for the sampling weights. The influential functions for the hazard ratios were used as the auxiliary variables to be calibrated and the proposed calibrated estimator was shown to effectively reduce the phase 2 variance component. \citet{lumley2011connections} calibrated the influential functions for the hazard ratios derived from an auxiliary model and gained substantial precisions. The auxiliary model is based on data in which the variable of interest, absent in the sample, is imputed through a Cox model that exclusively utilizes the true values within the sample. 

The aforementioned calibration procedure modifies weights for a given sample. The same result can also be obtained by taking an appropriate sample. Balanced sampling has a long history and finds application in diverse contexts, with a variety of proposed techniques \citep{tille2011ten}. It incorporates the same idea as calibration, which involves modifying the sampling weights to ensure that the Horvitz-Thompson estimator of the total of the auxiliary variables aligns with its corresponding population quantity. However, there is a fundamental difference between the two approaches: balanced sampling is primarily a sampling technique, whereas calibration is an estimation technique. After the introduction of random balanced sampling by \citet{yates1946review}, many methods have been proposed for the selection of random balanced samples. The most recent method for selecting a random balanced sample is the cube method proposed by \citet{deville2004efficient}; \citet{chauvet2006fast} further proposed a fast algorithm to implement the cube method in order to select a balanced sample; \citet{tille2004coordination} provides solutions for some extended problems regarding the practical implementation of the cube method; \citet{deville2005variance} derived a general approximation of variance under balanced sampling based on a residual technique; \citet{tille2011ten} provides a review and overview of the application of balanced sampling through the cube method. \citet{chauvet2011optimal} studied a method for computing optimal inclusion probabilities for balanced sampling on given auxiliary variables; \cite{breidt2012penalized} suggested a penalized balanced sampling method utilizing linear mixed models at the design stage to include auxiliary variables; \citet{jauslin2021enhanced} presented a method to handle the selection of a balanced sample in a highly stratified population; \citet{tille2022some} proposed an approach that combines the cube method with multivariate matching; \citet{ali2022balanced} proposed to use the cube method for selection of PSU’s in a two-stage design. Recently, balanced sampling has been actively used in the area of environmental sampling. \citet{stevens2004spatially} first introduced the term spatially balanced sampling and since then many efforts have been made in this area \citep{grafstrom2012spatially, grafstrom2013doubly, brown2015spatially, grafstrom2018spatially, kermorvant2019spatially, robertson2022spatially, xie2022spatially}. Despite its development in the aforementioned domains, to the best of our knowledge, there has not been an application of balanced sampling in the context of fitting a Cox model for survival times under a two-phase design. 

In this paper, we propose a balanced sampling procedure for selecting subjects under a stratified case-cohort design. Here, we consider the case-cohort design, in which all cases are selected with a sampling probability of 1. Naturally, we focus on developing a balanced sampling procedure for selecting a subcohort in the non-cases. The popular Cox model is postulated for the underlying failure times. The typical weighted estimating functions are then employed to estimate regression parameters in a Cox model.


The remainder of the paper is organized as follows. In Section~\ref{sec:back}, we present the fundamental concepts of the Cox model and case-cohort design. In Section~\ref{sec:bs}, we introduce the concept of balanced sampling, which plays a crucial role in improving the efficiency and accuracy of our study's design and analysis. 
In Section~\ref{sec:proposed}, we introduce the proposed method, which is implementing the balanced sampling in a stratified case-cohort design. In Section~\ref{sec:simulation}, we conduct extensive simulation studies to investigate and compare the performance of our proposed method and the existing method with a simple random sampling. In Section~\ref{sec:nwts}, we apply the proposed method to the National Wilms Tumor Study dataset to assess the performance of balanced sampling. Concluding remarks are given in Section~\ref{sec:diss}.

\section{Case-Cohort Design and Cox Model} \label{sec:back}

\subsection{Study Design and Data}

Consider a study cohort of $N$ subjects. 
Let $i = 1, \ldots, n$ 
denote the subject in the cohort.
Let $T_i$ and $C_i$ be the potential event time and the potential censoring time, respectively, where
the subscript $i$ is for the $i$th subject. 
$X_i=\min(T_i,C_i)$ denote the observed time and $\Delta_i=I(T_i < C_i)$ is the corresponding failure indicator.
Let $\mathbf{Z}_i$ be a vector of covariates. The observed data for the study cohort is then $\{X_i, \Delta_i, \mathbf{Z}_i\}, i=1, \ldots, N$.

\begin{figure}[htp]
    \centering
    \includegraphics[width=\textwidth]{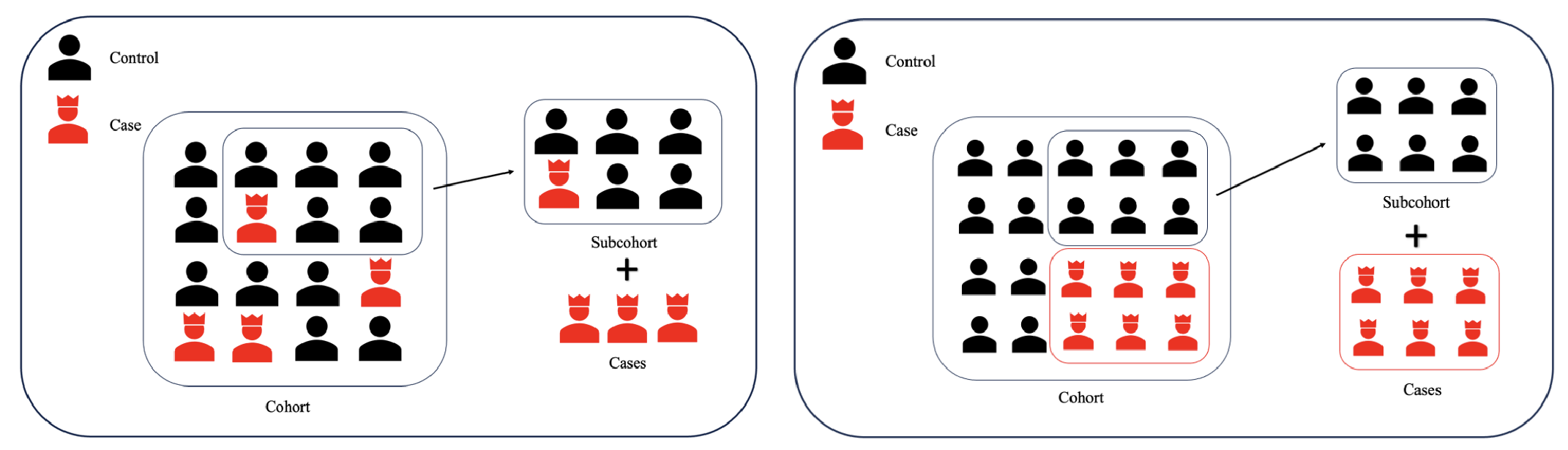}
    \caption{An illustration of a case-cohort design (left) and a stratified case-cohort design (right).}
    \label{fig:cch}
\end{figure}

Under a conventional case-cohort design, subjects are sampled disproportionately. 
Specifically, a subset of size $n_{sc}$, called subcohort, is sampled via simple random sampling. 
Then, cases, subjects experiencing the event of interest, who are not included in the subcohort are added to the subcohort.
Thus, a case-cohort sample of size $n$ is comprised of all cases and a random subset of non-cases with corresponding sampling probabilities being 1 and $p_{sc} (< 1)$, respectively. The case-cohort design can also be seen as a two-phase design, where in the first phase, the cohort itself can be viewed as a sample from an unknown superpopulation, and in the second phase, a case-cohort sample is selected. The left panel in Figure~\ref{fig:cch} is an illustration of the second phase of this design. 

This design can be extended to a stratified version whose stratification variables are based on those available for the entire cohort members \citep{borgan2000exposure,kulich2004improving}. 
Let $\mathbf{Z^{\ast}}$ be a vector of covariates available for the entire cohort member. 
Examples of $\mathbf{Z}^{\ast}$s include auxiliary variables that are both cost-effective and simple to measure, while also being closely correlated with the variable of interest. These auxiliary variables are accessible across the entire cohort.
The study cohort can be stratified based on some variables available for the entire cohort member. For example, when $\Delta_i$s $(i=1, \ldots, N)$ are used for stratification, the corresponding stratified case-cohort design can be implemented retrospectively. After ascertaining the failure status, subjects can be stratified into a stratum of cases and a stratum of non-cases. Sampling of subjects is then conducted within each stratum separately, as shown in the right panel of Figure~\ref{fig:cch}. This design can be further extended to accommodate a sampling of cases and/or further stratification of non-cases based on $\mathbf{Z}^{\ast}_i$s. The observed data for the corresponding case-cohort sample is $\{X_i, \Delta_i, \mathbf{Z}_i,  \mathbf{Z}^{\ast}_i\}, i=1, \ldots, n$. Note that for the cohort members not included in the case-cohort sample, one is assumed to observe $\{X_i, \Delta_i, \mathbf{Z}^{\ast}_i \}$ only.

\subsection{Model and Estimation} \label{sec:mod_est}


For the association between the potential failure time $T$ and a set of covariate $\mathbf{Z}$, we impose the popular Cox proportional hazards (PH) model \citep{cox1972regression}. For a given $\mathbf{Z}_i$, the Cox PH model for the hazard function at time $t$, $\lambda (t|\mathbf{Z}_i)$ is defined as  
\begin{equation} \label{mod_cph}
    \lambda (t|\mathbf{Z}_i) = \lambda_0(t)\exp(\bm{\beta ^{\top}} \mathbf{Z}_i)
\end{equation}
where $\lambda_0(\cdot)$ is the baseline hazard function, $\bm{\beta}$ is a vector of regression parameters, and ${\top}$ denotes the transpose of a vector or a matrix.
$\bm{\beta}$ in \eqref{mod_cph} can be estimated by maximizing the following partial likelihood function:
\begin{equation*}
    L(\bm{\beta})=\prod_{i=1}^N \left\{\frac{\exp(\bm{\beta}^{\top} \mathbf{Z}_{i})}{\sum_{l \in R_i} \exp(\bm{\beta}^{\top} \mathbf{Z}_{l})}\right\}^{\Delta_i}
\end{equation*}
where $R_i$ is the risk set at $i$th observed time. 
The maximum partial likelihood estimator for $\bm{\beta}$ can be equivalently obtained by finding the solution to the following estimating equation (\ref{equation:U}): 
    \begin{equation}
        \mathbf{U}(\bm{\beta}) = \sum_{i=1}^N \mathbf{U}_i(\bm{\beta}) = \sum_{i=1}^N \int_0^{\infty} \{\mathbf{Z_i}-\bar{\mathbf{Z}}(\bm{\beta},t) \}dN_i(t) = 0
        \label{equation:U}
    \end{equation}
    where 
    \begin{equation*}
       \bar{\mathbf{Z}}(\bm{\beta},t)=\frac{\sum_{i=1}^N Y_i(t) \mathbf{Z}_i\exp(\bm{\beta}^{\top}\mathbf{Z}_i)}{\sum_{i=1}^N Y_i(t)\exp(\bm{\beta}^{\top} \mathbf{Z}_i)},
    \end{equation*}
$N_i(t)=I(X_i \leq t, \Delta = 1)$ is the per subject counting process which counts the number of observed failures and $Y_i(t)=I(X_i \geq t)$ is the per subject risk process which takes value 1 at times at which the $i$th subject is at risk for failure and value 0 otherwise.

Under a case-cohort design, \eqref{equation:U} cannot be evaluated since $Z_i$s are available only for the subjects in the case-cohort sample. To deal with this missingness in $Z_i$s, a weighted estimating equations approach has typically been used \citep{prentice1986case,kulich2004improving}. The sampling weight of a subject, defined as the inverse of the sampling probability of a subject, is incorporated in \eqref{equation:U}. Let $\xi_i$ and $\pi_i$ be the sampling indicator and sampling probability of the $i$th subject, respectively. Then, the corresponding sampling weight for the $i$th subject is defined as $w_i = \xi_i/\pi_i$. The regression parameters $\beta$ in \eqref{mod_cph} can be estimated by solving the weighted estimating equations $U_{w}(\beta)=0$. Specifically,

\begin{equation} \label{eq: est_eq_in_cch}
        \mathbf{U}_{w}(\bm{\beta})=\sum_{i=1}^N \int_0^{\infty} w_i \{\mathbf{Z}_i-\bar{\mathbf{Z}}_{w}(\bm{\beta},t) \}dN_i(t) = 0
\end{equation}
where

\begin{equation*}
       \bar{\mathbf{Z}}_{w}(\bm{\beta},t)=\frac{\sum_{i=1}^N w_i Y_i(t)\mathbf{Z}_i\exp(\bm{\beta}^{\top}\mathbf{Z}_i)}{\sum_{i=1}^N w_i Y_i(t)\exp(\bm{\beta}^{\top}\mathbf{Z}_i)}, 
\end{equation*}

The values of $\pi_i$ are determined by the associated case-cohort designs. For example, under the traditional case-cohort design, $\pi_i = 1$ for the subjects who experienced the event of interest while $\pi_i = n_{c}/N_c$ where 
$n_{c}$ and $N_c$ denote the numbers of subjects in the subcohort and in the cohort who did not experience the event of interest, respectively \citep{lin1994semiparametric,barlow1999analysis}. Under a stratified case-cohort design, different $\pi_i$s are assigned for different strata \citep{borgan2000exposure,kulich2004improving}. By incorporating auxiliary information, $\pi_i$ can either be estimated or calibrated to enhance the efficiency of the resulting estimator \citep{breslow2009improved,breslow2009using}. The resulting estimators have been shown to be consistent and asymptotically normal whose finite asymptotic covariance function can be consistently estimated \citep{kulich2004improving, breslow2009improved}.

\section{Balanced Sampling} \label{sec:bs}

In this section, we briefly introduce the balanced sampling design and an algorithm that can implement this design.

\subsection{Balanced Sampling} \label{subsec:bs}
In the balanced sampling design, a random sample is chosen for the Horvitz-Thompson estimator of the total of auxiliary variables to be equal to their corresponding population totals \citep{deville2004efficient}. 
Specifically, we consider a sample $S$ of size $n$ from a finite population $U$ of size $N$ with each subject having an inclusion probability $\pi_i$ of being selected into the sample. Let $\mathbf{x}_i=(x_{i1}  \ldots, x_{ip})^{\top}$ denote the vector of auxiliary variables, where the $p$ auxiliary variables can be collected for all the subjects in the population. The balanced sampling design is then said to be balanced if the following balancing equation holds \citep{deville2004efficient}: $\sum_{i \in U} (x_{ij}/\pi_i)  u_i = \sum_{i \in U}x_{ij}$. Thus, the auxiliary variables are exactly estimated by the balanced sampling design without errors. 
The same property can also be obtained by calibrating sampling weights to the population total of the auxiliary variables. Note that this calibration is an estimation method for a given sample that adjusts sampling weights to ensure accurate estimates of auxiliary variables whereas the balanced sampling is a sampling approach that automatically selects a sample that satisfies this calibration property. The cube method \citep{deville2004efficient} can be employed to implement a balanced sampling design. 

\subsection{The Cube Method} \label{subsec:cube}

\begin{figure}[htp]
    \centering
    \includegraphics[width=0.6\textwidth]{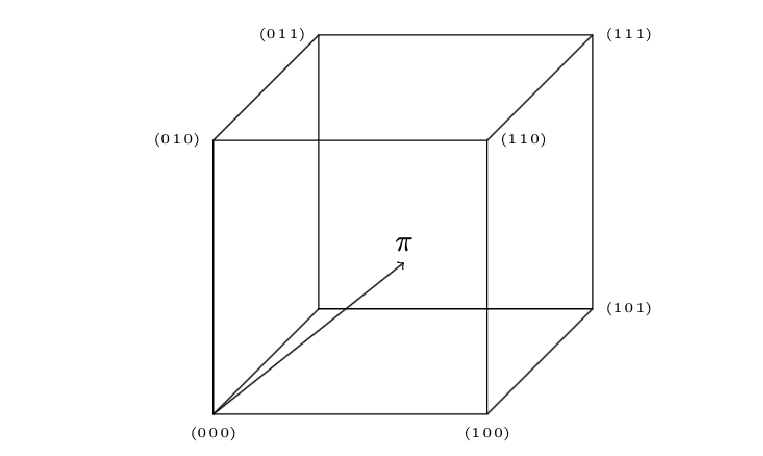}
    \caption{Illustration of the cube when the population size is $3$. $\pi$ is the inclusion probability vector.}
    \label{fig:cube_n3}
\end{figure}

The cube method \citep{deville2004efficient} is a geometric representation of the balanced sampling to select a balanced sample. The idea underlying this method is that samples can be represented using a vector of sampling indicators, which can be expressed as vertices of an $N$-cube. Figure~\ref{fig:cube_n3} is an illustration of the cube when the population size is $3$ and $\pi$ is the inclusion probability vector \citep{deville2004efficient}. A balancing subspace serves the role of constraining the conditions for achieving a balanced sample, and the cube method selects a vertex that lies at the intersection of the cube and the balancing subspace. The balancing subspace is defined as an affine subspace in $\mathbb{R}^N$ of dimension $N-p$ denoted by $Q$ where $Q=\left\{ \mathbf{u} \in \mathbb{R}^N | \sum_{i \in U} (\mathbf{x}_i/{\pi_i}) u_i = \sum_{i \in U} \mathbf{x}_i \right\}$ \citep{tille2011ten}.  The cube method consists of two phases: the flight phase and the landing phase. The flight phase starts as a random walk from the inclusion probabilities vector and continues within the intersection of the cube and balancing subspace. The random walk stops at one of the vertices within the intersection of the cube and the balancing subspace. When one cannot find an exactly balanced sample, the landing phase starts to function to produce an approximately balanced sample. The landing phase can be achieved either by a linear programming or a suppression of variables. The former and latter are more suitable for cases with small population sizes and large population sizes with a large number of auxiliary variables ($p > 20$). For more detailed information on the cube method, please refer to \citet{deville2004efficient}.

\section{Proposed Method} \label{sec:proposed}

In this section, we introduce our proposed balanced sampling design to implement in selecting a subcohort under a case-cohort design. Specifically, we consider a stratified case-cohort design stratified by the failure status, where all the cases are selected with a sampling probability of 1, and a subcohort consisting of non-cases (controls) is sampled using the proposed balanced sampling design instead of the traditional simple random sampling. Note that this design can be easily extended to accommodate a stratified sampling of controls based on some variables available for the entire cohort members. 
As stated in Section~\ref{sec:mod_est}, we assume a Cox PH model~\eqref{mod_cph} for the relationship between $\mathbf{Z}$ and $T$. Furthermore, we assume that a vector of auxiliary variables $\mathbf{Z}^{\ast}_i, i = 1, \ldots, N$, highly correlated with the covariates of interest $\mathbf{Z}_i, i = 1, \ldots, N$ are available for the entire cohort members.

Conducting a balanced sampling from the study cohort in the context of fitting a Cox PH model can be viewed as a calibration problem of a population total defined as the sum of $\mathbf{U}_i(\bm{\beta})$s $i=1, \ldots, N$ in~\eqref{equation:U} \citep{breslow2009improved,breslow2009using,lumley2011connections}. Note that, as pointed out by \citet{lumley2011connections}, $\mathbf{Z}^{\ast}$ is not suitable to use as the auxiliary variables for calibration of the sum of $\mathbf{U}_i(\bm{\beta})$s. Auxiliary variables typically employed in this context have been so-called delta-betas, denoted as $\mathbf{\Delta} \bm{\beta}_{(i)}$. The delta-beta ($\mathbf{\Delta} \bm{\beta}_{(i)}$) represents the influence of the $i$th case on the estimation of the coefficient $\bm{\beta}$ and can be approximated by $\mathbf{\Delta} \bm{\beta}_{(i)} \approx \mathbfcal{I}^{-1}(\bm{\hat{\beta}}) \mathbf{U}_i(\bm{\hat{\beta}})$,  where $\bm{\hat{\beta}}$ is the estimated regression coefficient and $\mathbfcal{I}^{-1}(\bm{\hat{\beta}})$ is the inverse of the information matrix. $\mathbf{\Delta} \bm{\beta}_{(i)}$ exhibits a strong correlation with $\mathbf{U}_i(\bm{\hat{\beta}})$, making it a good candidate for the auxiliary variables in the balanced sampling procedure, as discussed in \cite{lumley2011connections}.
Thus, we adapt this idea and first calculate delta-betas based on $\mathbf{Z}^{\ast}_i$s by fitting a Cox PH model on $\mathbf{Z}^{\ast}_i$s. The next step is to implement the aforementioned cube method to the stratum of controls to select a balanced sample balancing on both $\mathbf{\Delta} \bm{\beta}_{(i)}$ and $\pi_i$. Here, $\pi_i$ serves to determine the balanced sample size \citep{deville2004efficient}, and therefore, we include it for completeness. 
For the landing phase of the cube method, we employ the aforementioned linear programming to produce an approximate balanced sample. To expedite the sampling procedure, we implement the fast flight algorithm \citep{chauvet2006fast} in the flight phase.


The estimator of $\bm{\beta}$ is defined as the solution to \eqref{eq: est_eq_in_cch} 
and denoted by $\bm{\hat{\beta}}_w$. For variance estimation, we proposed to use a robust sandwich-type estimator that adapts a simplified approximation considered by \citet{hajek1981sampling} and \citet{deville2005variance} for a balanced sample:
\begin{equation}
    \text{Var}(\bm{\hat{\beta}}_w) = \mathbfcal{I}^{-1}(\bm{\hat{\beta}}_w) + \mathbfcal{I}^{-1}(\bm{\hat{\beta}}_w) V(y) \mathbfcal{I}^{-1}(\bm{\hat{\beta}}_w), \label{eq: var_est}
\end{equation}
where 
\begin{equation*}
    V(y) = \sum_{i \in S} c_i \frac{\left\{\mathbf{y}_i - \mathbf{x}_i^{\top}(\sum_{l \in S} c_l \frac{\mathbf{x}_l \mathbf{x}_l^{\top}}{\pi_l^2})^{-1} \sum_{l \in S} c_l \frac{\mathbf{x}_l \mathbf{y}_l}{\pi_l}\right\}^{\otimes 2}}{\pi_i^2},
    \label{eq: var_est}
\end{equation*}
,$c_i = \frac{n}{n-p} (1-\pi_i)$, $\mathbf{y}_i=\mathbf{U}_i(\bm{\beta})$, $\mathbf{x}_i=\mathbf{\Delta} \bm{\beta}_{(i)}$, and $\mathbf{a}^{\otimes 2} = \mathbf{aa}^{\top}$. The variance estimator comprises two components: $\mathbfcal{I}^{-1}(\bm{\hat{\beta}}_w)$ estimates the phase one variability due to sampling of the cohort from an unknown super population, and $\mathbfcal{I}^{-1}(\bm{\hat{\beta}}_w) V(y) \mathbfcal{I}^{-1}(\bm{\hat{\beta}}_w)$ estimates the phase two variability due to sampling the case-cohort sample from the given cohort.
Note that $c_i$ approximates $d_i$ defined as the solution of the following nonlinear system:
 \begin{equation*}
    \label{Eq:Var_Appro_nonlinear_sys}
    \pi_i(1-\pi_i)=d_i - \frac{d_i \mathbf{x}_i^{\top}}{\pi_i} \left[ \sum_{l=1}^N d_l \frac{\mathbf{x}_l \mathbf{x}_l^{\top}}{\pi_l^2} \right]^{-1} \frac{d_i \mathbf{x}_i}{\pi_i}, \ i=1, \ldots, N.
\end{equation*}

The asymptotic property of the estimator $\hat{\beta}_w$ mirrors that of the estimator derived from calibrated weights, as previously demonstrated in \citet{breslow2009improved}. The implementation of the proposed balanced sampling procedure is summarized in Algorithm 1.


\begin{algorithm} 
\label{algo: algo1}
\caption{Implementation of the Proposed Method}
\textbf{Input:} Cohort data with size $N$; \\
\textbf{Output:} $\bm{\hat{\beta}}_w$ and $\text{SE} = \sqrt{\text{Var}(\bm{\hat{\beta}}_w)}$;

\textbf{Step 1.} Fit an auxiliary Cox model with auxiliary variables $\mathbf{x}_i$ and get delta-beta $\mathbf{x}_i=\mathbf{\Delta} \bm{\beta}_{(i)}$ for balanced sampling;\\
\textbf{Step 2.} Choose a cohort and define the index set of the full cohort $S=\{ 1,2,..., N \}, i \in S$;\\
\textbf{Step 3.} Implement the cube method in the control set $S_C=S_{\delta=0}$ with size $N_c$ and select a subcohort of size $n_c$ with auxiliary variable $x_i$ and inclusion probability $\pi_i=n_c/N_c$, denote the index set of the subcohort as $S_{BS}$;\\
\textbf{Step 4.} Fit a Cox model using the case-cohort sample $S_{CCS}=S_F \cup S_{BS}$ where $S_F=S_{\delta=1}$ is the index set of events and estimate the hazard ratio $\bm{\hat{\beta}}_w$ and standard error $\text{SE}$;

\Return{$\bm{\hat{\beta}}_w$ and $\text{SE}$}\;

\end{algorithm}

\section{Simulation Studies} \label{sec:simulation}

In this section, we investigate and evaluate the finite sample performance of the proposed design under various types of covariates, cohort sizes, sample sizes, and censoring proportions. Our investigation is centered around two distinct cohort setups: fixed cohorts and random cohorts. In the fixed cohorts configuration, we can explore the unbiasedness and variability arising from samplings within a cohort,  phase 2 of the two-phase design. On the other hand, in the random cohort setting, we investigate both the variabilities of sampling a cohort and case-cohort samples within the cohort, namely, phase 1 and phase 2 variabilities. In each case, we focus on two different censoring proportions: 20\% and 90\%. This allows us to evaluate the effectiveness of balanced sampling in a low-censoring setting over the simple random sampling design and verify its suitability for the case-cohort design in a high-censoring setting where its application is most appropriate.

\subsection{Simulation Setup I} \label{subsec:sim_1}
We first generate the failure time $T$ and generate the censoring time $C$ from an exponential distribution with a constant rate being adjusted to achieve the desired censoring proportions of $20\%$ and $90\%$. 
Two covariates $ \mathbf{Z}=(Z_1, Z_2)^{\top}$ are generated from a bivariate normal distribution with mean $\bm{\mu} = (0, 0)^{\top}$ and a correlation coefficient of $0.8$ and $0.5$ between the two covariates. 
The true value of the Cox model coefficient corresponding to $Z_1$ is set to $\beta = \log 2$ and the coefficient for $Z_2$ is set to 0. For the cohort sizes, we consider $N=1000$ and $N=3000$. For $N=1000$, we consider two sample sizes, namely subcohort sizes, $n=100$ and $n=200$. For $N=3000$, we consider two sample sizes $n=300$, and $n=600$. To generate binary covariates, we assign a value of 1 to the bivariate normal covariates that are greater than 0 and 0 to the rest. To implement the cube method, we use the \texttt{samplecube} function in the \texttt{sampling} package \citep{tille2012survey} in \texttt{R}, which implements the fast flight algorithm to select a balanced sample. To fit a Cox model, we use the \texttt{coxph} function in \texttt{survival} \citep{therneau2015package} package. For each configuration, we repeatedly select a subcohort $2000$ times using both a simple random sampling and a balanced sampling. A Cox model is then fitted to each sample. The mean and standard deviation of $2000$ estimated regression coefficients are calculated to evaluate the performance of two sampling designs, where the standard deviation represents the phase 2 variability of this design. The phase 2 variability is estimated using the proposed robust-sandwich estimator (\ref{eq: var_est}).

\begin{figure}
    \includegraphics[width=\textwidth]{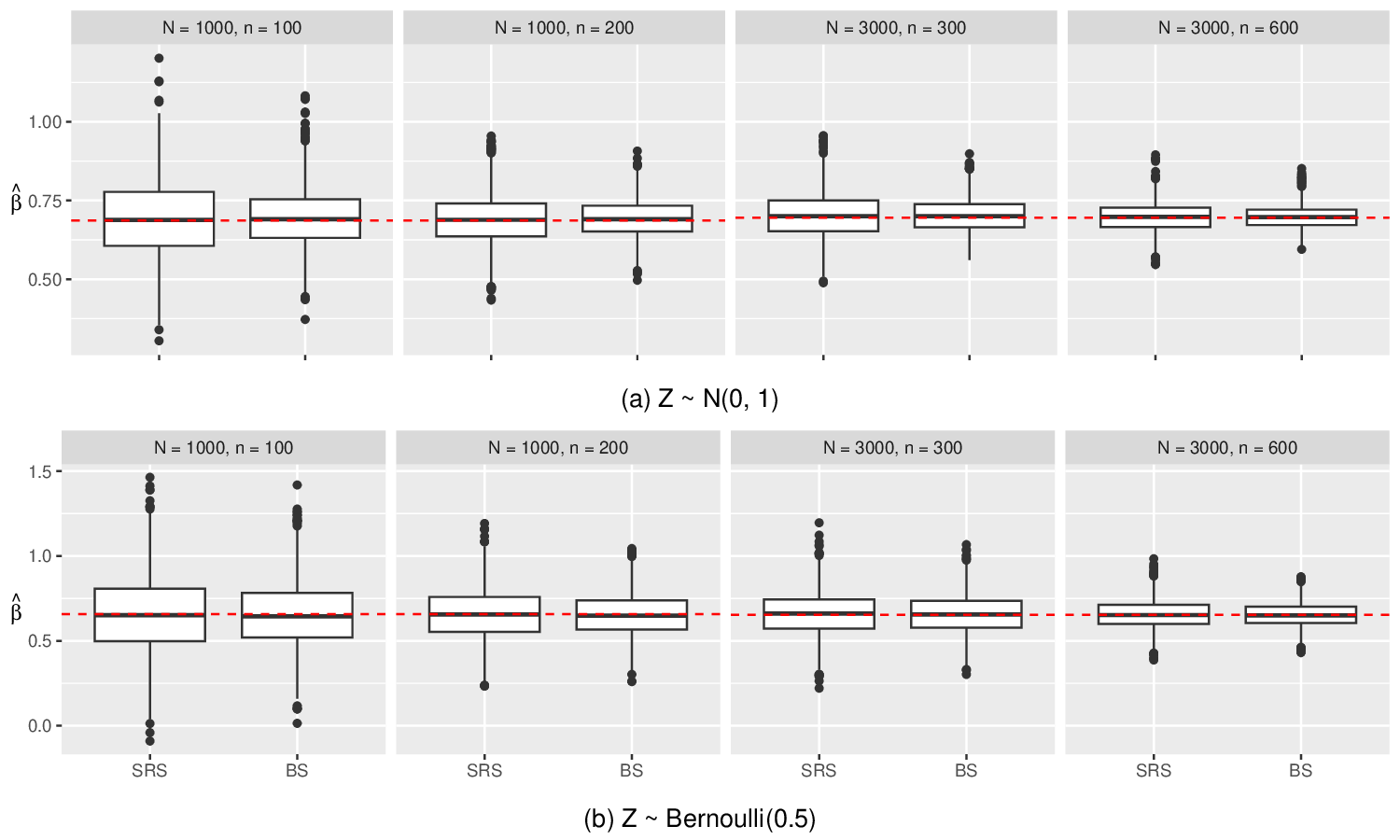}
    \caption{Boxplots of estimated regression coefficients for a fixed cohort based on subcohort samplings from a fixed cohort with the censoring proportion of $20\%$ and correlation between covariates being $\rho=0.8$ for Simulation Setup I. The red dotted horizontal line represents the estimated regression coefficient based on the full cohort data. (a) refers to the results for continuous covariates and (b) refers to the results for binary covariates. $N$ denotes the cohort size, $n$ denotes the subcohort size, and $\hat{\beta}$ denotes the estimated regression coefficients. SRS and BS denote simple random sampling and balanced sampling, respectively.}
    \label{fig:fix_low_censor}
\end{figure}

Simulation results under a low censoring rate of $20\%$ are displayed in Figure~\ref{fig:fix_low_censor}. Under the setup considered with varying sizes of the full cohort, sample, and types of covariates, the variabilities of the estimated coefficients under the balanced sampling design are significantly smaller compared with those under the simple random sampling counterpart. The point estimates are approximately unbiased and the variance estimation is highly accurate (Table \ref{tab: low_censor} in \hyperref[sec:appendix]{Appendix}). This demonstrates that the proposed balanced sampling outperforms simple random sampling with a higher efficiency. Under a high-censoring proportion of 90\%, we additionally consider a stratified case-cohort sampling. For comparison purposes, we also perform two types of calibrations: calibration (CAL) to the case-cohort sample with a simple random sampling \citep{breslow2009improved} and calibration conducted after a balanced sampling (BSc). Note that calibration and balanced sampling provide theoretically identical results. \cite{tille2011ten} suggested that re-calibrating auxiliary variables during the estimation stage after conducting balanced sampling may enhance efficiency. Therefore, we aim to investigate the consistency between the balanced sampling and calibration and a potential improvement by applying a calibration after the balanced sampling.

Figure~\ref{fig: fix_0.8} and Table~\ref{tab: high_censor_0.8} in \hyperref[sec:appendix]{Appendix} show that, in both subcohort sampling and stratified case-cohort design, balanced samples result in an improved estimation, along with a reduction in variabilities, and the standard error estimation is accurate. The relative efficiency for each method is calculated and it is defined as the ratio between the standard deviation of the 2000 Cox model estimates of the given design and the standard error of the Cox model within the full cohort context. A lower relative efficiency value indicates a more efficient design. As indicated in Table~\ref{tab: high_censor_0.8}, the efficiency of the balanced sampling method enhanced in comparison to simple random sampling. When a subcohort of size 100 is selected from a cohort of size 1000 with the binary covariates, approximately $3.3\% \sim 4\%$ of the $2000$ samples are removed from the calculation of the simulation results due to the perfect separation issues, i.e. either only 0s or 1s are observed for the cases in the sample. Calibration and balanced sampling produce similar results while the re-calibration after balanced samplings reduces the variability of the estimated coefficients, but the improvement is minimal. The results for the setting with a moderate correlation of $0.5$ are available in Table~\ref{tab: high_censor_0.5} and Figure~\ref{fig: fix_0.5} in \hyperref[sec:appendix]{Appendix}. Although the results remain similar, it is evident that the variability reduction is considerably smaller compared to the case with the stronger correlation of $0.8$.

\begin{figure}
    \centering
    \includegraphics[width=\textwidth]{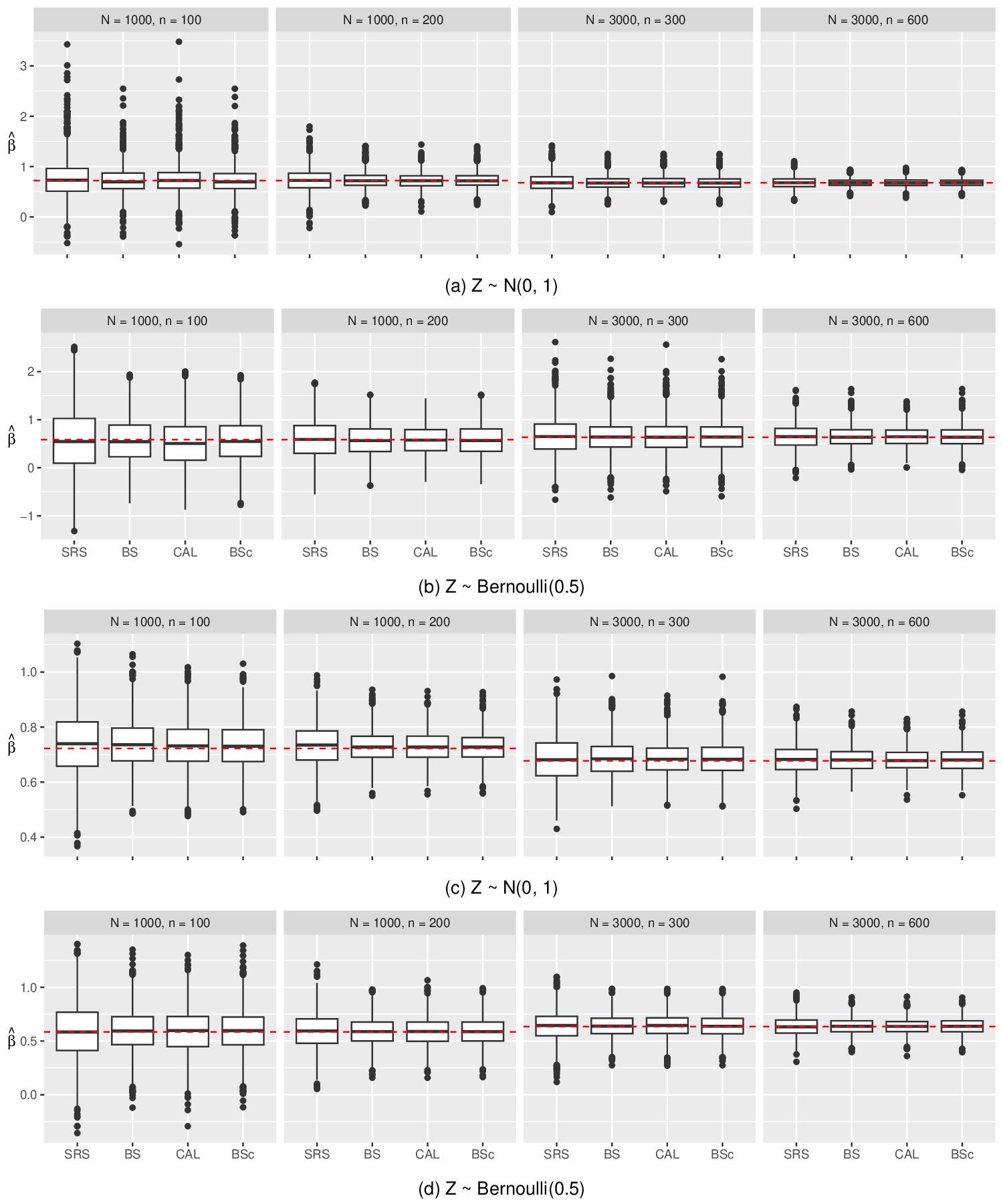}
    \caption{Boxplots of estimated regression coefficients for a fixed cohort with the censoring proportion of $90\%$ and correlation between covariates being $\rho=0.8$ for Simulation Setup I. The red-dotted horizontal line represents the estimated regression coefficient based on the full cohort data. (a) refers to the results from subcohort samplings for continuous covariates, (b) refers to the results from subcohort samplings for binary covariates, (c) refers to the results from case-cohort samplings for continuous covariates, and (d) refers to the results from case-cohort samplings for binary covariates. $N$ denotes the cohort size, $n$ denotes the subcohort size, and $\hat{\beta}$ denotes the estimated regression coefficients. SRS, BS, CAL, and BSc denote simple random sampling, balanced sampling, calibration, and re-calibration after balanced sampling, respectively.}
    \label{fig: fix_0.8}
\end{figure}

\subsection{Simulation Setup II}

In this setting, we aim to consider the variability of generating a study cohort in addition to the variability of generating a case-cohort sample within the study cohort. Namely, we consider both the phase 1 variability and phase 2 variability in a two-phase design. The settings for generating cohort data remain the same as those used in Simulation Setup \hyperref[subsec:sim_1]{I} except that, during $2000$ iterations, a new cohort is generated for each iteration. Similarly, we investigate both the setups with the low and high censoring proportions with $20\%$ and $90\%$. We first generate the $2000$ cohort data and only perform a subcohort sampling within each cohort for the low censoring proportion scenario. Simulation results are shown in Figure \ref{fig:random_low_censor} and Table~\ref{tab: c1s1_low_censor} in Appendix. In this case, the findings are similar to those of the fixed cohort case in \hyperref[subsec:sim_1]{I}. Under the balanced sampling design, the variability of the estimated coefficients decreases compared to that under the simple random sampling design. The variances of the resulting estimator are accurately estimated with the total estimated variance being approximately equal to the sum of phase 1 and phase 2 components.

For the high censoring proportion setting with $90\%$, the simulation results are available in Figure~\ref{fig: random_0.8} and Table~\ref{tab: c1s1_high_censor_0.8} in Appendix. As in Simulation Setup \hyperref[subsec:sim_1]{I} with the binary covariate and $n=100$, unstable estimates are produced due to the perfect separation issue and, thus, are excluded from the calculation. Such problematic samples, however, only constitute $3\% \sim 3.5\%$ of $2000$ samples. Similar to the fixed cohort case, in both subcohort sampling and stratified case-cohort design, the proposed balanced sampling produces estimates with reduced variabilities compared with those under the simple random sampling counterpart. This can also be further confirmed by the reduction in relative efficiency, indicating enhanced performance. As the cohort size and the subcohort size increase, the estimated standard errors, the summation of the phase 1 and the phase 2 standard errors, get closer to the true variabilities. The proposed estimated coefficients are close to their true values. The performance of the proposed balanced sampling and the calibration methods exhibit minimal differences, and incorporating re-calibration after balanced sampling results in slight enhancements in the reduction of variability. The simulation results of the setup with the weaker correlation of $0.5$ are provided in Table~\ref{tab: c1s1_high_censor_0.5} and Figure \ref{fig: random_0.5} in \hyperref[sec:appendix]{Appendix}. Similarly, in the setting with a correlation of $0.5$ between the covariates, the proposed balanced sampling still perform better in almost all configuration considered but the reduction in variability is smaller when compared to the case with the strong correlation of $0.8$.

\begin{figure}
    \centering
    \includegraphics[width=\textwidth]{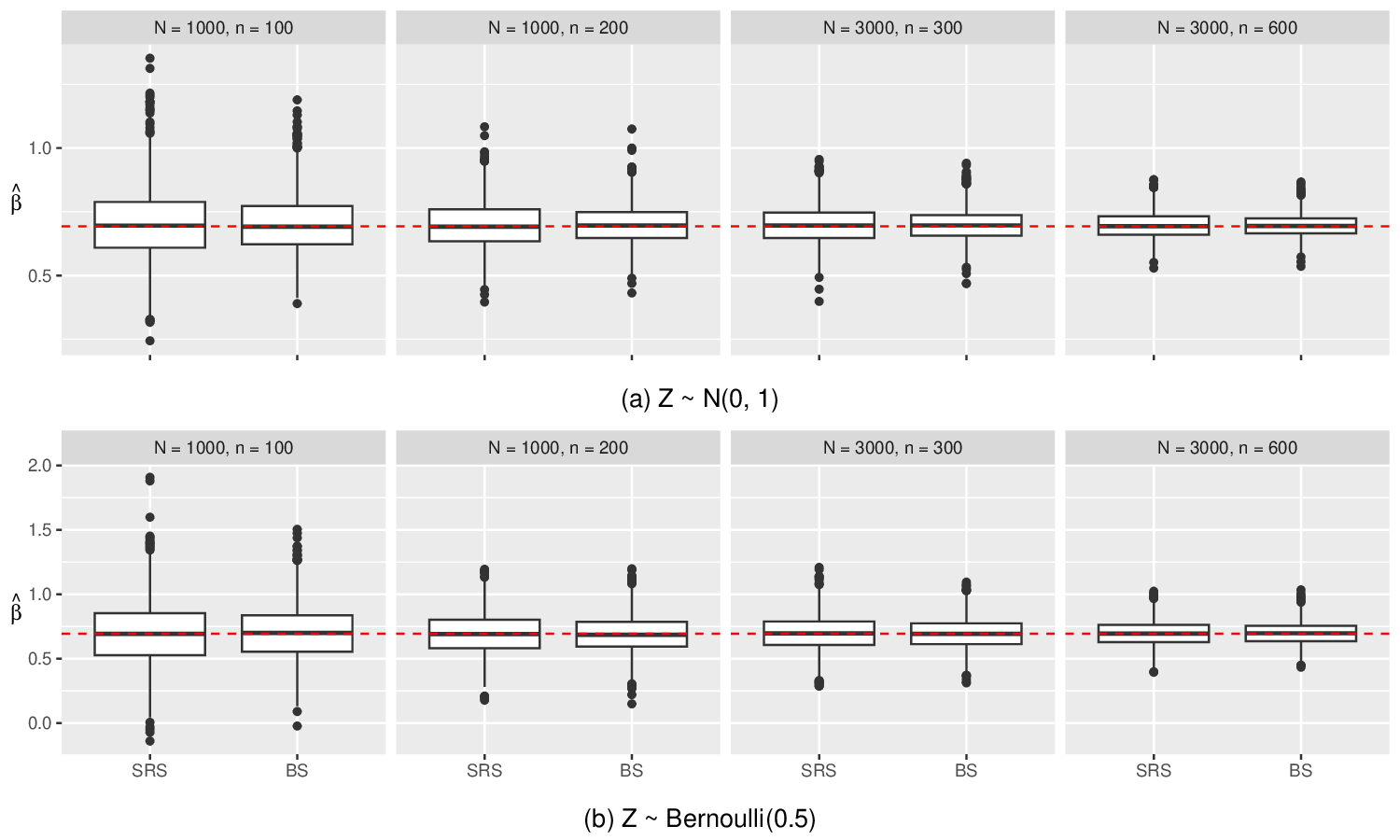}
    \caption{Boxplots of estimated regression coefficients for random cohorts based on subcohort samplings from random cohorts with the censoring proportion of $20\%$ and correlation between covariates being $\rho=0.8$ for Simulation Setup II. The red dotted horizontal line represents the estimated regression coefficient based on the full cohort data, $\log{2}$. (a) refers to the results for continuous covariates and (b) refers to the results for binary covariates. $N$ denotes the cohort size, $n$ denotes the subcohort size, and $\hat{\beta}$ denotes the estimated regression coefficients. SRS and BS denote simple random sampling and balanced sampling, respectively.}
    \label{fig:random_low_censor}
\end{figure}

\begin{figure}
    \centering
    \includegraphics[width=\textwidth]{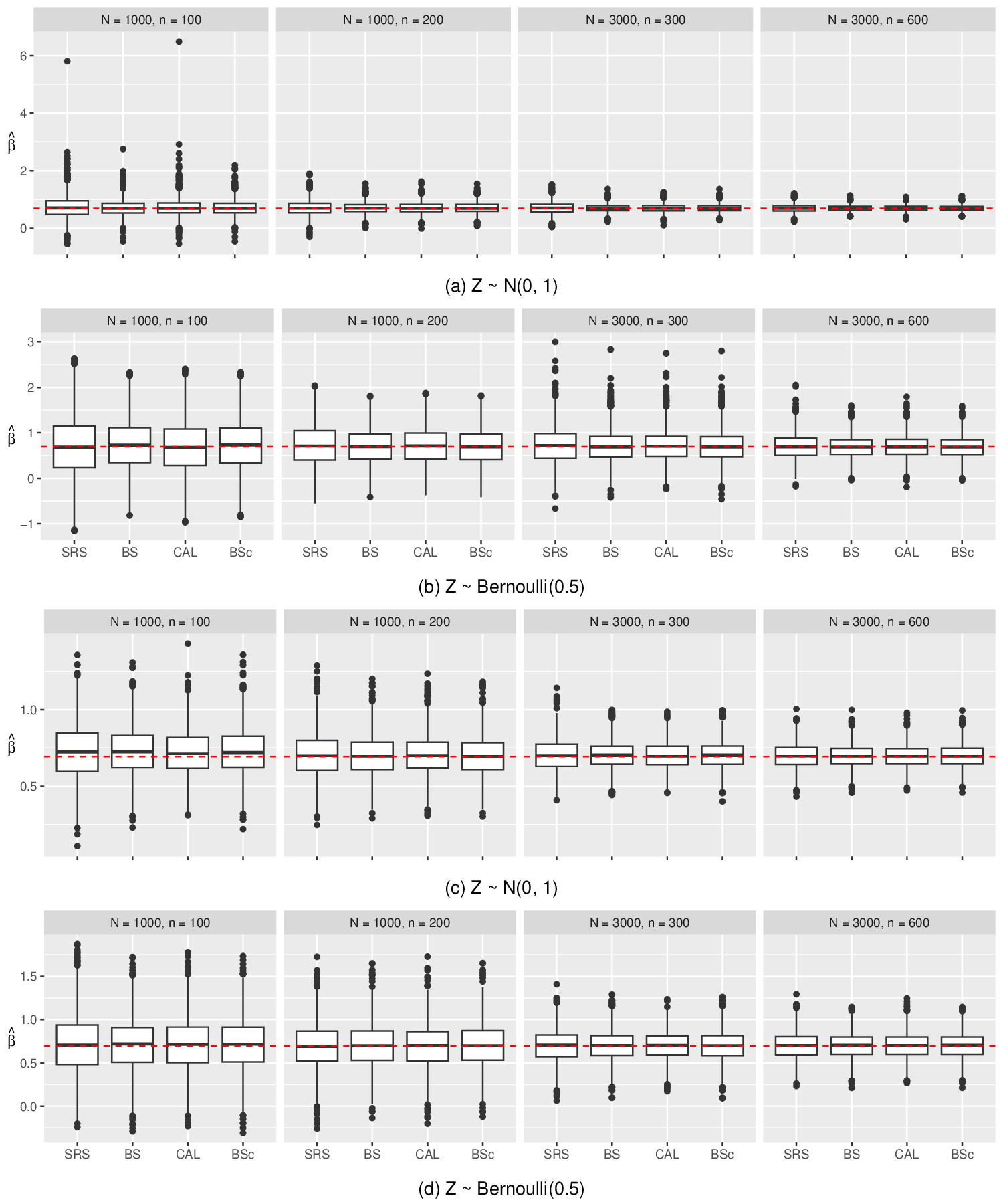}
    \caption{Boxplots of estimated regression coefficients for random cohorts with the censoring proportion of $90\%$ and correlation between covariates being $\rho=0.8$ for Simulation Setup II. The red-dotted horizontal line represents the estimated regression coefficient based on the full cohort data, $\log{2}$. (a) refers to the results from subcohort samplings for continuous covariates, (b) refers to the results from subcohort samplings for binary covariates, (c) refers to the results from case-cohort samplings for continuous covariates, and (d) refers to the results from case-cohort samplings for binary covariates. $N$ denotes the cohort size, $n$ denotes the subcohort size, and $\hat{\beta}$ denotes the estimated regression coefficients. SRS, BS, CAL, and BSc denote simple random sampling, balanced sampling, calibration, and re-calibration after balanced sampling, respectively. }
    \label{fig: random_0.8}
\end{figure}

\section{Application to National Wilms Tumor Study} \label{sec:nwts}

The National Wilms Tumor Study (NWTS) was the initial clinical research unit dedicated to pediatric intergroup studies in North America during 1980-1994 \citep{d1989treatment, green1998comparison}. The study was designed to enhance the survival rates of children diagnosed with Wilms tumor and it encompasses valuable data and insights regarding pediatric patients with this condition. The NWTS cohort consists of $3915$ patients and it is available in the \texttt{addhazard} package \citep{hu1addhazard} in \texttt{R}. Event-free survival, stage of disease, histology evaluated at a registering institution (unfavorable vs favorable), histology evaluated by the central reference laboratory (unfavorable vs favorable), age at diagnosis, and the diameter of tumor are available.

We implement a stratified case-cohort design, which was also considered in \citet{breslow2009using,breslow2009improved,kulich2004improving}.  
Sixteen strata are constructed based on the failure status,
stage (I, II vs III, IV), institutional histology (unfavorable vs favorable), and age group (age $<$ 1 vs age $\geq$ 1). All subjects from the smallest $13$ strata are included in the sample, resulting in a sampling weight of $1$ for each of these subjects. We further select sample sizes $120$, $160$, and $120$ from the three largest strata, resulting in a phase 2 sample of $669$ cases and $648$ sampled controls. More details on the stratified sampling design can be found in Table~\ref{tab: StratiDes} in Appendix. We select samples from the three strata using simple random sampling and the proposed balanced sampling. We compare the performance of the two sampling methods by assessing the variability of the estimated hazard ratios. To assess the performance of calibration in comparison to balanced sampling, we additionally conduct separate calibration analyses and re-calibration after applying the balanced sampling technique. We repeat the sampling process $2000$ times.

As the underlying model for the true failure time, we consider a Cox model.  
The Cox model included the central histology (UH: 0 = Favorable, 1 = Unfavorable), age as a piecewise linear term with a change point at one year (Age1: 1 = stage I, II, Age0: 0 = stage III, IV),  stage indicator (Stage), tumor diameter (Diameter), the interaction term between age and histology (UH $\times$ Age1, UH $\times$ Age0), and the interaction term between stage and tumor diameter (Stage $\times$ Diameter) as covariates. We constructed the delta-betas using an auxiliary model that incorporated histology obtained from registering institutions. This model also considered age as a piecewise linear term with a change point at one year (Age1: 1 = stage I, II, Age0: 0 = stage III, IV), a stage indicator (Stage), tumor diameter (Diameter), as well as interaction terms between age and histology (UH $\times$ Age1, UH $\times$ Age0) and between stage and tumor diameter (Stage $\times$ Diameter) as covariates. These covariates were consistent with those used in the main model of interest. We estimate the regression parameters using the samples selected by a simple random sampling within each stratum (SRS), and samples selected by the proposed balanced sampling (BS). For a comparison purpose, we calibrate weights for the samples selected by a simple random sampling (CAL) and re-calibrate weights after selecting balanced samples (BSc). 

Results are summarized in Figure~\ref{fig:nwts_stratification} and Table~\ref{tab: nwts_phase2_stratification} in \hyperref[sec:appendix]{Appendix}. 
BS significantly improved precision for all variables except for the histology and age interaction term. The averaged estimated coefficients of BS are all very close to the estimates using the full cohort data. This can also be confirmed by the relative efficiency values. Some point estimates, however, including UH, Diameter, UH, and age interactions under SRS are somewhat different from the full cohort estimates. In general, the results of BS, CAL, and BSc are close to each other, as evidenced in the simulation experiments. In summary, the results suggest that our proposed balanced sampling procedure and relevant estimation method effectively reduce variability and provide reliable estimates of the regression parameters in a Cox model. Performing a calibration for the samples selected by the proposed balanced sampling procedure does not seem to improve precision.

\begin{figure}
    \centering
    \includegraphics[width=\textwidth]{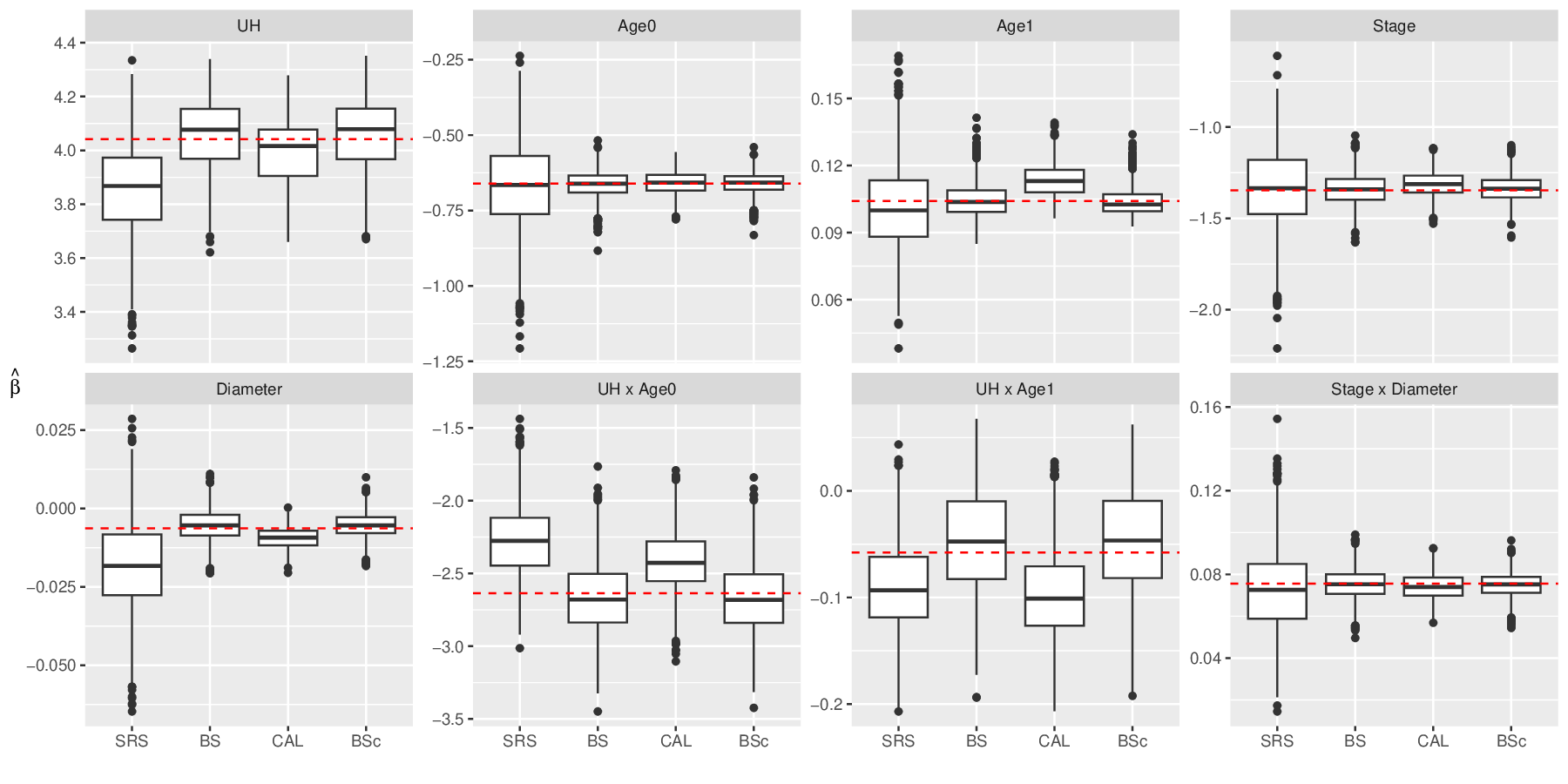}
    \caption{Boxplots of the regression coefficients estimates based on a stratified sampling design for NWTS. The red-dotted horizontal line represents the estimated regression coefficient based on the full cohort data. $\hat{\beta}$ denotes the Cox model coefficients. SRS, BS, CAL, and BSc denote simple random sampling, balanced sampling, calibration, and re-calibration after balanced sampling, respectively. }
    \label{fig:nwts_stratification}
\end{figure}

\section{Discussion and Conclusions} \label{sec:diss}

The case-cohort design serves as an excellent alternative when limited resources prevent access to complete cohort data or complete information of interest. Typically, a subcohort selection involves a simple random sampling or stratified simple random sampling. To obtain a more efficient sample, we propose to adopt a balanced sampling procedure that utilizes auxiliary information in selecting the subcohort when fitting a Cox regression model. This approach aims to optimize the selection process and improve the overall efficiency of the resulting estimator.

The balanced sampling procedure guarantees that the auxiliary variable, which exhibits a high correlation with the variable of interest and is available at the full cohort level, is estimated without errors. This approach enhances the accuracy of estimating the variable of interest by incorporating its relationship with the auxiliary variable. It can also be viewed as a sampling design that achieves the calibration of the auxiliary variables while a calibration modifies the sampling weights to achieve this property for a given sample. 
Both approaches aim to ensure that the selected sample accurately estimates the population quantities of interest for auxiliary information. We use the cube method \citep{deville2004efficient} to implement the proposed balanced sampling procedure.

The performances of the proposed balanced sampling procedure and resulting estimator are verified through extensive simulation experiments. Compared to the results under a simple random sampling, the proposed balanced sampling yields samples with reduced variability for the estimators of the regression parameters of a Cox model. Calibrating the sampling weights under a simple random sampling produces similar results to those under our proposed balanced sampling procedure, as expected. Additionally, the proposed method demonstrates efficient results when applied to the NWTS data, further validating its effectiveness and practicability.

The proposed procedure can be naturally extended to accommodate a generalized case-cohort design that allows a sampling of cases \cite{Kang:Cai:marg:2009} and/or clustered failure times \citep{Kang:Cai:margin:2009,Kang:Cai:marg:2009}.

\section*{Code Availability} \label{sec:computation}
This paper presents the results obtained using version 4.3.1 of the statistical computing environment R. A sample code to reproduce the results can be found in \texttt{https://github.com/kchoi59/CchBs/tree/main}.

\section*{Acknowledgements}

    {
    \bibliographystyle{statcomp}
    \bibliography{bibliography}
    }

\clearpage

\setcounter{section}{1}
\renewcommand*{\thesection}{\Alph{section}}
\section*{Appendix} \label{sec:appendix}


\begin{table}[!hp]\centering
\caption{Simulation results for Simulation Setup I: samplings from a fixed cohort with the censoring proportion of $20\%$ and correlation between covariates being $\rho=0.8$. FC, SRS, and BS denote the full cohort, simple random sampling, and balanced sampling, respectively. CO and SUBCO denote the cohort and subcohort, respectively. $N$ denotes the cohort size, $n$ denotes the subcohort size. $\hat{\beta}$ denotes the estimated regression coefficients. Mean denotes the mean of the estimated regression coefficients, SD denotes the standard deviation of the estimated regression coefficients, SE represents the mean of the estimated standard errors, and RE denotes the relative efficiency.}\label{tab: low_censor}
\scriptsize
\begin{tabular*}{\textwidth}{@{\extracolsep\fill}ccrrrrrrrrr@{}}\toprule
& &\multicolumn{4}{c}{Continuous Covariates} &\multicolumn{4}{c}{Binary Covariates} \\\cmidrule(lr){3-6} \cmidrule(lr){7-10} 
&CO &\multicolumn{2}{c}{$N = 1000$} &\multicolumn{2}{c}{$N = 3000$} &\multicolumn{2}{c}{$N = 1000$} &\multicolumn{2}{c}{$N = 3000$} \\\cmidrule(lr){3-4}  \cmidrule(lr){5-6} \cmidrule(lr){7-8} \cmidrule(lr){9-10}
&SUBCO &$n = 100$ &$n = 200$ &$n = 300$ &$n = 600$ &$n = 100$ &$n = 200$ &$n = 300$ &$n = 600$ \\\midrule
\multirow[t]{2}{*}{FC} &$\hat{\beta}$ &\multicolumn{2}{c}{0.6864} &\multicolumn{2}{c}{0.6954} &\multicolumn{2}{c}{0.6571} &\multicolumn{2}{c}{0.6530} \\
&SE &\multicolumn{2}{c}{0.0410} &\multicolumn{2}{c}{0.0234} &\multicolumn{2}{c}{0.0732} &\multicolumn{2}{c}{0.0420} \\
\multirow[t]{2}{*}{SRS} &Mean &0.6941 &0.6901 &0.7029 &0.6976 &0.6548 &	0.6552&	0.6606	&0.6549 \\
&SD &0.1246 &0.0816 &0.0716 &0.0473 &0.2235	&0.1488&0.1284	&0.0860 \\
&RE &3.0390	&1.9902	&3.0598	&2.0214	&3.0533	&2.0328	&3.0571	&2.0476 \\
\multirow[t]{2}{*}{BS} &Mean &0.6952 &0.6934 &0.7026 &0.6977 &0.6536	&0.6558&	0.6579	&0.6533 \\
&SD &0.0951 &0.0613 &0.0548 &0.0370 &0.1948	&0.1263	&0.1141	&0.0731 \\
&SE &0.0915 &0.0605 &0.0539 &0.0357 &0.1794&	0.1208&	0.1081	&0.0726 \\
&RE &2.3195	&1.4951	&2.3419	&1.5812	&2.6612	&1.7254	&2.7167	&1.7405 \\

\bottomrule
\end{tabular*}
\end{table}

\clearpage
\begin{table}[!htp]\centering
\caption{Simulation results for Simulation Setup I: samplings from a fixed cohort with the censoring proportion of $90\%$ and correlation between covariates being $\rho=0.8$. FC, SC, and CCS denote the full cohort, subcohort sampling, and stratified case-cohort design, respectively. CO and SUBCO denote the cohort and subcohort, respectively. $N$ denotes the cohort size, $n$ denotes the subcohort size. SRS, BS, CAL, and BSc denote simple random sampling, balanced sampling, calibration, and re-calibration after balanced sampling, respectively. $\hat{\beta}$ denotes the estimated regression coefficients. Mean denotes the mean of the estimated regression coefficients. SD denotes the standard deviation of the estimated regression coefficients, SE represents the mean of the estimated standard errors, and RE denotes the relative efficiency.}\label{tab: high_censor_0.8}
\scriptsize
\begin{tabular*}{\textwidth}{@{\extracolsep\fill}cccrrrrrrrrr@{}}\toprule
\multirow{5}{*}{} &\multirow{3}{*}{} & &\multicolumn{4}{c}{Continuous Covariates} &\multicolumn{4}{c}{Binary Covariates} \\\cmidrule(lr){4-7} \cmidrule(lr){8-11} 
& &CO &\multicolumn{2}{c}{$N = 1000$} &\multicolumn{2}{c}{$N = 3000$} &\multicolumn{2}{c}{$N = 1000$} &\multicolumn{2}{c}{$N = 3000$} \\\cmidrule(lr){4-5}  \cmidrule(lr){6-7} \cmidrule(lr){8-9} \cmidrule(lr){10-11}
& &SUBCO &$n = 100$ &$n = 200$ &$n = 300$ &$n = 600$ &$n = 100$ &$n = 200$ &$n = 300$ &$n = 600$ \\\midrule
\multirow[t]{2}{*}{FC} & &$\hat{\beta}$ &\multicolumn{2}{c}{0.7220} &\multicolumn{2}{c}{0.6772} &\multicolumn{2}{c}{0.5852} &\multicolumn{2}{c}{0.6351} \\
& &SE &\multicolumn{2}{c}{0.0993} &\multicolumn{2}{c}{0.0575} &\multicolumn{2}{c}{0.2112} &\multicolumn{2}{c}{0.1222} \\
& & & & & & & & & & \\
\multirow[t]{10}{*}{SC} &\multirow[t]{2}{*}{SRS} &Mean &0.7558 &0.7302 &0.6857 &0.6792 &0.5784 &0.5953 &0.6605 &0.6484 \\
& &SD &0.3887 &0.2248 &0.1759 &0.1128 &0.6959 &0.4247 &0.4052 &0.2520 \\
& &RE &3.9144	&2.2638	&3.0591	&1.9617&3.2950	&2.0109	&3.3159	&2.0622 \\
&\multirow[t]{4}{*}{BS} &Mean &0.7250 &0.7289 &0.6788 &0.6792 &0.5624 &0.5714 &0.6491 &0.6483 \\
& &SD &0.2675 &0.1537 &0.1253 &0.0754 &0.4985 &0.3387 &0.3211 &0.2142 \\
& &SE &0.2164 &0.1394 &0.1167 &0.0762 &0.4929 &0.3228 &0.3072 &0.2029 \\
& &RE &2.6939	&1.5478	&2.1791	&1.3113	&2.3603	&1.6037	&2.6277	&1.7529 \\
&\multirow[t]{2}{*}{CAL} &Mean &0.7431 &0.7254 &0.6816 &0.6800 &0.5300 &0.5778 &0.6510 &0.6502 \\
& &SD &0.2928 &0.1526 &0.1286 &0.0785 &0.5214 &0.3186 &0.3370 &0.2110 \\
& &RE &2.9486	&1.5368	&2.2365	&1.3652	&2.4688	&1.5085	&2.7578	&1.7267\\
&\multirow[t]{2}{*}{BSc} &Mean &0.7220 &0.7277 &0.6783 &0.6793 &0.5612 &0.5722 &0.6499 &0.6481 \\
& &SD &0.2624 &0.1509 &0.1244 &0.0752 &0.4921 &0.3374 &0.3203 &0.2140 \\
& &RE & 2.6425	&1.5196	&2.1635	&1.3078	&2.3300	&1.5975	&2.6211	&1.7512\\
& & & & & & & & & & \\
\multirow[t]{10}{*}{CCS} &\multirow[t]{2}{*}{SRS} &Mean &0.7388 &0.7326 &0.6860 &0.6829 &0.5912 &0.5930 &0.6402 &0.6344 \\
& &SD &0.1185 &0.0770 &0.0850 &0.0545 &0.2665 &0.1685 &0.1377 &0.0895 \\
& &RE &1.1934	&0.7754	&1.4783	&0.9478	&1.2618	&0.7978	&1.1268	&0.7324 \\
&\multirow[t]{4}{*}{BS} &Mean &0.7390 &0.7294 &0.6875 &0.6815 &0.5956 &0.5887 &0.6395 &0.6363 \\
& &SD &0.0875 &0.0567 &0.0647 &0.0426 &0.2034 &0.1315 &0.1041 &0.0725 \\
& &SE &0.0727 &0.0508 &0.0546 &0.0383 &0.1772 &0.1223 &0.1025 &0.0684 \\
& &RE &0.8812	&0.5710	&1.1252	&0.7409	&0.9631	&0.6226	&0.8519	&0.5933 \\
&\multirow[t]{2}{*}{CAL} &Mean &0.7348 &0.7301 &0.6856 &0.6805 &0.5908 &0.5884 &0.6430 &0.6359 \\
& &SD &0.0854 &0.0551 &0.0616 &0.0412 &0.2051 &0.1314 &0.1068 &0.0709 \\
& &RE & 0.8600	&0.5549	&1.0713	&0.7165	&0.9711	&0.6222	&0.8740	&0.5802\\
&\multirow[t]{2}{*}{BSc} &Mean &0.7335 &0.7278 &0.6866 &0.6811 &0.5956 &0.5882 &0.6393 &0.6362 \\
& &SD &0.0822 &0.0542 &0.0627 &0.0418 &0.2022 &0.1307 &0.1037 &0.0724 \\
& &RE &0.8278	&0.5458	&1.0904	&0.7270	&0.9574	&0.6188	&0.8486	&0.5925 \\
\bottomrule
\end{tabular*}
\end{table}

\newpage
\begin{table}[!htp]\centering
\caption{Simulation results for Simulation Setup I: samplings from a fixed cohort with the censoring proportion of $90\%$ and correlation between covariates being $\rho=0.5$. FC, SC, and CCS denote the full cohort, subcohort sampling, and stratified case-cohort design, respectively. CO and SUBCO denote the cohort and subcohort, respectively. $N$ denotes the cohort size, $n$ denotes the subcohort size. SRS, BS, CAL, and BSc denote simple random sampling, balanced sampling, calibration, and re-calibration after balanced sampling, respectively. $\hat{\beta}$ denotes the estimated regression coefficients. Mean denotes the mean of the estimated regression coefficients. SD denotes the standard deviation of the estimated regression coefficients, SE represents the mean of the estimated standard errors, and RE denotes the relative efficiency.}\label{tab: high_censor_0.5}
\scriptsize
\begin{tabular*}{\textwidth}{@{\extracolsep\fill}cccrrrrrrrrr@{}}\toprule
\multirow{5}{*}{} &\multirow{3}{*}{} & &\multicolumn{4}{c}{Continuous Covariates} &\multicolumn{4}{c}{Binary Covariates} \\\cmidrule(lr){4-7} \cmidrule(lr){8-11} 
& &CO &\multicolumn{2}{c}{$N = 1000$} &\multicolumn{2}{c}{$N = 3000$} &\multicolumn{2}{c}{$N = 1000$} &\multicolumn{2}{c}{$N = 3000$} \\\cmidrule(lr){4-5}  \cmidrule(lr){6-7} \cmidrule(lr){8-9} \cmidrule(lr){10-11}
& &SUBCO &$n = 100$ &$n = 200$ &$n = 300$ &$n = 600$ &$n = 100$ &$n = 200$ &$n = 300$ &$n = 600$ \\\midrule
\multirow[t]{2}{*}{FC} & &$\hat{\beta}$ &\multicolumn{2}{c}{0.6413} &\multicolumn{2}{c}{0.7784} &\multicolumn{2}{c}{0.6509} &\multicolumn{2}{c}{0.6199} \\
& &SE &\multicolumn{2}{c}{0.1070} &\multicolumn{2}{c}{0.0590} &\multicolumn{2}{c}{0.2095} &\multicolumn{2}{c}{0.1207} \\
& & & & & & & & & & \\
\multirow[t]{10}{*}{SC} &\multirow[t]{2}{*}{SRS} &Mean &0.6617 &0.6549 &0.7706 &0.7758 &0.6396 &0.6588 &0.6418 &0.6304 \\
& &SD &0.4013 &0.2419 &0.1880 &0.1220 &0.6768 &0.4168 &0.3911 &0.2488 \\
& &RE & 3.7505	&2.2607	&3.1864	&2.0678	&3.2305	&1.9895	&3.2403	&2.0613\\
&\multirow[t]{4}{*}{BS} &Mean &0.6509 &0.6432 &0.7744 &0.7735 &0.6335 &0.6423 &0.6429 &0.6274 \\
& &SD &0.3595 &0.2036 &0.1750 &0.1067 &0.6344 &0.4075 &0.3786 &0.2358 \\
& &SE &0.2887 &0.1867 &0.1643 &0.1081 &0.6204 &0.4038 &0.3533 &0.2312 \\
& &RE &3.3598	&1.9028	&2.9661	&1.8085	&3.0282	&1.9451	&3.1367	&1.9536 \\
&\multirow[t]{2}{*}{CAL} &Mean &0.6659 &0.6550 &0.7718 &0.7757 &0.6404 &0.6594 &0.6412 &0.6317 \\
& &SD &0.3718 &0.2096 &0.1763 &0.1127 &0.6841 &0.4025 &0.3781 &0.2390 \\
& &RE & 3.4748	&1.9589	&2.9881	&1.9102	&3.2654	&1.9212	&3.1326	&1.9801\\
&\multirow[t]{2}{*}{BSc} &Mean &0.6520 &0.6434 &0.7743 &0.7735 &0.6343 &0.6412 &0.6431 &0.6276 \\
& &SD &0.3555 &0.2022 &0.1748 &0.1066 &0.6373 &0.4064 &0.3788 &0.2358 \\
& &RE &3.3224	&1.8897	&2.9627	&1.8068	&3.0420	&1.9399	&3.1384	&1.9536 \\
& & & & & & & & & & \\
\multirow[t]{10}{*}{CCS} &\multirow{2}{*}{SRS} &Mean &0.6560 &0.6489 &0.7860 &0.7827 &0.6657 &0.6616 &0.6236 &0.6196 \\
& &SD &0.1185 &0.0773 &0.0821 &0.0565 &0.2517 &0.1677 &0.1369 &0.0888 \\
& &RE & 1.1075	&0.7224	&1.3915	&0.9576	&1.2014	&0.8005	&1.1342	&0.7357\\
&\multirow[t]{4}{*}{BS} &Mean &0.6542 &0.6482 &0.7866 &0.7834 &0.6489 &0.6555 &0.6226 &0.6212 \\
& &SD &0.1138 &0.0700 &0.0775 &0.0509 &0.2310 &0.1539 &0.1278 &0.0821 \\
& &SE &0.0968 &0.0679 &0.0722 &0.0497 &0.2121 &0.1447 &0.1206 &0.0803 \\
& &RE & 1.0636	&0.6542	&1.3136	&0.8627	&1.1026	&0.7346&	1.0588	&0.6802\\
&\multirow[t]{2}{*}{CAL} &Mean &0.6495 &0.6455 &0.7852 &0.7831 &0.6635 &0.6605 &0.6223 &0.6197 \\
& &SD &0.1138 &0.0733 &0.0782 &0.0536 &0.2309 &0.1551 &0.1250 &0.0814 \\
& &RE & 1.0636	&0.6850	&1.3254	&0.9085	&1.1021	&0.7403	&1.0356	&0.6744\\
&\multirow[t]{2}{*}{BSc} &Mean &0.6536 &0.6479 &0.7866 &0.7834 &0.6487 &0.6557 &0.6225 &0.6212 \\
& &SD &0.1149 &0.0703 &0.0774 &0.0509 &0.2306 &0.1536 &0.1278 &0.0821 \\
& &RE & 1.0738	&0.6570	&1.3119	&0.8627	&1.1007	&0.7332	&1.0588	&0.6802\\
\bottomrule
\end{tabular*}
 
\end{table}

\newpage
\begin{table}[!htp]\centering
\caption{Simulation results for Simulation Setup II: samplings from random cohorts with the censoring proportion of $20\%$ and correlation between covariates being $\rho=0.8$. FC, SRS, and BS denote the full cohort, simple random sampling, and balanced sampling, respectively. CO and SUBCO denote the cohort and subcohort, respectively. $N$ denotes the cohort size, $n$ denotes the subcohort size. $\hat{\beta}$ denotes the estimated regression coefficients. Mean denotes the mean of the estimated regression coefficients. SD denotes the standard deviation of the estimated regression coefficients. $\text{SE}_1$ and $\text{SE}_2$ denote the mean of phase 1 standard errors, and the mean of phase 2 standard errors, respectively. SE represents the mean of the estimated standard errors, and RE denotes the relative efficiency.}\label{tab: c1s1_low_censor}
\scriptsize
\begin{tabular*}{\textwidth}{@{\extracolsep\fill}ccrrrrrrrrr@{}}\toprule
& &\multicolumn{4}{c}{Continuous Covariates} &\multicolumn{4}{c}{Binary Covariates} \\\cmidrule(lr){3-6} \cmidrule(lr){7-10} 
&CO &\multicolumn{2}{c}{$N = 1000$} &\multicolumn{2}{c}{$N = 3000$} &\multicolumn{2}{c}{$N = 1000$} &\multicolumn{2}{c}{$N = 3000$} \\\cmidrule(lr){3-4}  \cmidrule(lr){5-6} \cmidrule(lr){7-8} \cmidrule(lr){9-10}
&SUBCO &$n = 100$ &$n = 200$ &$n = 300$ &$n = 600$ &$n = 100$ &$n = 200$ &$n = 300$ &$n = 600$ \\\midrule
\multirow[t]{2}{*}{FC} &Mean &0.6933 &0.6943 &0.6938 &0.6934 &0.6925 &0.6919 &0.6938 &0.6948 \\
&SD &0.0408 &0.0402 &0.0238 &0.0234 &0.0722 &0.0739 &0.0438 &0.0426 \\
\multirow[t]{2}{*}{SRS} &Mean &0.7021 &0.6971 &0.6985 &0.6960 &0.6963 &0.6913 &0.6981 &0.6964 \\
&SD &0.1391 &0.0933 &0.0766 &0.0536 &0.2446 &0.1634 &0.1390 &0.0960 \\
&RE &3.4093	&2.3209	&3.2185	&2.2906	&3.3878	&2.2111	&3.1735	&2.2535 \\
\multirow[t]{6}{*}{BS} &Mean &0.7005 &0.6981 &0.6979 &0.6953 &0.7027 &0.6892 &0.6940 &0.6967 \\
&SD &0.1130 &0.0757 &0.0622 &0.0450 &0.2110 &0.1462 &0.1204 &0.0878 \\
&$\text{SE}_1$ &0.0424 &0.0417 &0.0239 &0.0237 &0.0755 &0.0745 &0.0429 &0.0427 \\
&$\text{SE}_2$ &0.0950 &0.0629 &0.0545 &0.0365 &0.1881 &0.1249 &0.1085 &0.0725 \\
&SE &0.1042 &0.0756 &0.0595 &0.0436 &0.2029 &0.1456 &0.1167 &0.0842 \\
&RE & 2.7696	&1.8831	&2.6134	&1.9231	&2.9224	&1.9783	&2.7489	&2.0610\\
\bottomrule
\end{tabular*}
\end{table}

\newpage
\begin{table}[!htp]\centering
\caption{Simulation results for Simulation Setup II: samplings from random cohorts with the censoring proportion of $90\%$ and correlation between covariates being $\rho=0.8$. FC, SRS, and BS denote the full cohort, simple random sampling, and balanced sampling, respectively. CO and SUBCO denote the cohort and subcohort, respectively. $N$ denotes the cohort size, $n$ denotes the subcohort size. $\hat{\beta}$ denotes the estimated regression coefficients. Mean denotes the mean of the estimated regression coefficients. SD denotes the standard deviation of the estimated regression coefficients. $\text{SE}_1$ and $\text{SE}_2$ denote the mean of phase 1 standard errors, and the mean of phase 2 standard errors, respectively. SE represents the mean of the estimated standard errors, and RE denotes the relative efficiency.}\label{tab: c1s1_high_censor_0.8}
\scriptsize
\begin{tabular*}{\textwidth}{@{\extracolsep\fill}cccccccccccc@{}}\toprule
& & &\multicolumn{4}{c}{Continuous Covariates} &\multicolumn{4}{c}{Binary Covariates} \\\cmidrule(lr){4-7} \cmidrule(lr){8-11} 
& &CO &\multicolumn{2}{c}{$N = 1000$} &\multicolumn{2}{c}{$N = 3000$} &\multicolumn{2}{c}{$N = 1000$} &\multicolumn{2}{c}{$N = 3000$} \\\cmidrule(lr){4-5}  \cmidrule(lr){6-7} \cmidrule(lr){8-9} \cmidrule(lr){10-11}
& &SUBCO &$n = 100$ &$n = 200$ &$n = 300$ &$n = 600$&$n = 100 $&$n = 200$ &$n = 300$ &$n = 600$ \\\midrule
\multirow[t]{14}{*}{SC} &\multirow[t]{2}{*}{FC} &Mean &0.6933 &0.6963 &0.6955 &0.6937 &0.6986 &0.7049 &0.6953 &0.6920 \\
& &SD &0.1037 &0.1002 &0.0587 &0.0583 &0.2093 &0.2165 &0.1191 &0.1193 \\
&\multirow[t]{2}{*}{SRS} &Mean &0.7362 &0.7103 &0.7084 &0.6980 &0.7015 &0.7297 &0.7284 &0.6997 \\
& &SD &0.4114 &0.2487 &0.2023 &0.1374 &0.7018 &0.4755 &0.4058 &0.2800 \\
& &RE & 3.9672	&2.4820	&3.4463	&2.3568	&3.3531	&2.1963	&3.4072	&2.3470\\
&\multirow[t]{6}{*}{BS} &Mean &0.7139 &0.7095 &0.7012 &0.6952 &0.7398 &0.7041 &0.7076 &0.6898 \\
& &SD &0.2824 &0.1860 &0.1363 &0.1006 &0.5755 &0.4075 &0.3437 &0.2359 \\
& &$\text{SE}_1$ &0.1145 &0.1082 &0.0615 &0.0600 &0.2310 &0.2189 &0.1246 &0.1217 \\
& &$\text{SE}_2$ &0.2179 &0.1386 &0.1193 &0.0779 &0.5307 &0.3447 &0.2997 &0.1973 \\
& &SE &0.2471 &0.1762 &0.1344 &0.0984 &0.5812 &0.4094 &0.3249 &0.2320 \\
& &RE &2.7232	&1.8563	&2.3220	&1.7256	&2.7496	&1.8822	&2.8858	&1.9774 \\
&\multirow[t]{2}{*}{CAL} &Mean &0.7249 &0.7082 &0.7029 &0.6978 &0.6960 &0.7216 &0.7154 &0.6980 \\
& &SD &0.3325 &0.1867 &0.1421 &0.1016 &0.5989 &0.4166 &0.3370 &0.2432 \\
& &RE & 3.2064	&1.8633	&2.4208	&1.7427	&2.8614	&1.9242	&2.8296	&2.0386\\
&\multirow[t]{2}{*}{BSc} &Mean &0.7124 &0.7101 &0.7009 &0.6950 &0.7392 &0.7019 &0.7072 &0.6900 \\
& &SD &0.2739 &0.1838 &0.1351 &0.1004 &0.5737 &0.4074 &0.3426 &0.2358 \\
& &RE &2.6413	&1.8343	&2.3015	&1.7221	&2.7410	&1.8818	&2.8766	&1.9765 \\
& & & & && & & & & \\
\multirow[t]{14}{*}{CCS} &\multirow[t]{2}{*}{FC} &Mean &0.6971 &0.6904 &0.6924 &0.6940 &0.6958 &0.6911 &0.6913 &0.6966 \\
& &SD &0.1038 &0.1048 &0.0598 &0.0587 &0.2195 &0.2108 &0.1186 &0.1224 \\
&\multirow[t]{2}{*}{SRS} &Mean &0.7281 &0.7047 &0.7032 &0.6961 &0.7148 &0.6963 &0.7010 &0.7006 \\
& &SD &0.1796 &0.1435 &0.1028 &0.0822 &0.3421 &0.2703 &0.1851 &0.1537 \\
& &RE & 1.7303	&1.3693	&1.7191	&1.4003	&1.5585	&1.2823	&1.5607	&1.2557\\
&\multirow[t]{6}{*}{BS} &Mean &0.7274 &0.7010 &0.7041 &0.6982 &0.7120 &0.7014 &0.6998 &0.6995 \\
& &SD &0.1531 &0.1309 &0.0856 &0.0730 &0.3031 &0.2466 &0.1713 &0.1449 \\
& &$\text{SE}_1$ &0.1060 &0.1047 &0.0598 &0.0595 &0.2131 &0.2113 &0.1210 &0.1205 \\
& &$\text{SE}_2$ &0.0840 &0.0608 &0.0548 &0.0380 &0.1818 &0.1235 &0.1090 &0.0730 \\
& &SE &0.1360 &0.1216 &0.0814 &0.0707 &0.2811 &0.2452 &0.1630 &0.1410 \\
& &RE & 1.4750	&1.2490	&1.4314	&1.2436	&1.3809	&1.1698	&1.4444	&1.1838\\
&\multirow[t]{2}{*}{CAL} &Mean &0.7194 &0.7039 &0.7006 &0.6966 &0.7145 &0.6985 &0.7014 &0.7001 \\
& &RE & 1.4355	&1.2214	&1.4448	&1.2283	&1.3731	&1.1883	&1.3820	&1.1863\\
& &SD &0.1490 &0.1280 &0.0864 &0.0721 &0.3014 &0.2505 &0.1639 &0.1452 \\
&\multirow[t]{2}{*}{BSc} &Mean &0.7262 &0.6995 &0.7034 &0.6983 &0.7126 &0.7018 &0.6996 &0.6994 \\
& &SD &0.1517 &0.1286 &0.0847 &0.0729 &0.3020 &0.2465 &0.1711 &0.1448 \\
& &RE & 1.4615	&1.2271	&1.4164	&1.2419	&1.3759	&1.1694	&1.4427	&1.1830\\
\bottomrule
\end{tabular*}
\end{table}

\newpage
\begin{table}[!htp]\centering
\caption{Simulation results for Simulation Setup II: samplings from random cohorts with the censoring proportion of $90\%$ and correlation between covariates being $\rho=0.5$. FC, SRS, and BS denote the full cohort, simple random sampling, and balanced sampling, respectively. CO and SUBCO denote the cohort and subcohort, respectively. $N$ denotes the cohort size, $n$ denotes the subcohort size. $\hat{\beta}$ denotes the estimated regression coefficients. Mean denotes the mean of the estimated regression coefficients. SD denotes the standard deviation of the estimated regression coefficients. $\text{SE}_1$ and $\text{SE}_2$ denote the mean of phase 1 standard errors, and the mean of phase 2 standard errors, respectively. SE represents the mean of the estimated standard errors, and RE denotes the relative efficiency.}\label{tab: c1s1_high_censor_0.5}
\scriptsize
\begin{tabular*}{\textwidth}{@{\extracolsep\fill}cccrrrrrrrrr@{}}\toprule
& & &\multicolumn{4}{c}{Continuous Covariates} &\multicolumn{4}{c}{Binary Covariates} \\\cmidrule(lr){4-7} \cmidrule(lr){8-11} 
& &CO &\multicolumn{2}{c}{$N = 1000$} &\multicolumn{2}{c}{$N = 3000$} &\multicolumn{2}{c}{$N = 1000$} &\multicolumn{2}{c}{$N = 3000$} \\\cmidrule(lr){4-5}  \cmidrule(lr){6-7} \cmidrule(lr){8-9} \cmidrule(lr){10-11}
& &SUBCO &$n = 100$ &$n = 200$ &$n = 300$ &$n = 600$ &$n = 100$ &$n = 200$ &$n = 300$ &$n = 600$ \\\midrule
\multirow[t]{14}{*}{SC} &\multirow[t]{2}{*}{FC} &Mean &0.6934 &0.6957 &0.6959 &0.6936 &0.7011 &0.7044 &0.6954 &0.6928 \\
& &SD &0.1035 &0.1008 &0.0587 &0.0583 &0.2137 &0.2148 &0.1182 &0.1205 \\
&\multirow[t]{2}{*}{SRS} &Mean &0.7317 &0.7113 &0.7062 &0.7015 &0.7067 &0.7419 &0.7309 &0.6984 \\
& &SD &0.1816 &0.1461 &0.1003 &0.0810 &0.6924 &0.4708 &0.4156 &0.2810 \\
& &RE & 1.7546	&1.4494	&1.7087	&1.3894	&3.2401	&2.1918	&3.5161	&2.3320 \\
&\multirow[t]{6}{*}{BS} &Mean &0.7247 &0.7076 &0.7065 &0.7026 &0.7261 &0.7435 &0.7111 &0.7052 \\
& &SD &0.1750 &0.1342 &0.0979 &0.0784 &0.6438 &0.4578 &0.3751 &0.2636 \\
& &$\text{SE}_1$ &0.1060 &0.1050 &0.0599 &0.0595 &0.2319 &0.2207 &0.1246 &0.1222 \\
& &$\text{SE}_2$ &0.1089 &0.0785 &0.0702 &0.0488 &0.6168 &0.4055 &0.3468 &0.2291 \\
& &SE &0.1527 &0.1315 &0.0926 &0.0771 &0.6596 &0.4620 &0.3686 &0.2597 \\
& &RE & 1.6908	&1.3313	&1.6678	&1.3448	&3.0126	&2.1313	&3.1734	&2.1876\\
&\multirow[t]{2}{*}{CAL} &Mean &0.7294 &0.7091 &0.7058 &0.7007 &0.7019 &0.7410 &0.7264 &0.6978 \\
& &SD &0.1732 &0.1412 &0.0963 &0.0786 &0.6842 &0.4621 &0.3937 &0.2704 \\
& &RE & 1.6734	&1.4008	&1.6405	&1.3482	&3.2017	&2.1513	&3.3308	&2.2440\\
&\multirow[t]{2}{*}{BSc} &Mean &0.7245 &0.7074 &0.7065 &0.7026 &0.7273 &0.7425 &0.7113 &0.7054 \\
& &SD &0.1742 &0.1344 &0.0977 &0.0783 &0.6418 &0.4559 &0.3756 &0.2634 \\
& &RE & 1.6831	&1.3333	&1.6644	&1.3431	&3.0033	&2.1224	&3.1777	&2.1859\\
& & & & && & & & & \\
\multirow[t]{14}{*}{CCS} &\multirow[t]{2}{*}{FC} &Mean &0.6948 &0.6933 &0.6937 &0.6948 &0.7077 &0.6998 &0.6952 &0.6960 \\
& &SD &0.1048 &0.1011 &0.0580 &0.0583 &0.2111 &0.2077 &0.1193 &0.1191 \\
&\multirow[t]{2}{*}{SRS} &Mean &0.7317 &0.7113 &0.7062 &0.7015 &0.7269 &0.7065 &0.7053 &0.7005 \\
& &SD &0.1816 &0.1461 &0.1003 &0.0810 &0.3262 &0.2601 &0.1850 &0.1549 \\
& &RE &1.7328	&1.4451	&1.7293	&1.3894	&1.5452	&1.2523	&1.5507	&1.3006 \\
&\multirow[t]{6}{*}{BS} &Mean &0.7247 &0.7076 &0.7065 &0.7026 &0.7352 &0.7070 &0.7052 &0.7008 \\
& &SD &0.1750 &0.1342 &0.0979 &0.0784 &0.3183 &0.2578 &0.1807 &0.1473 \\
& &$\text{SE}_1$ &0.1060 &0.1050 &0.0599 &0.0595 &0.2133 &0.2113 &0.1209 &0.1206 \\
& &$\text{SE}_2$ &0.1089 &0.0785 &0.0702 &0.0488 &0.2138 &0.1447 &0.1275 &0.0853 \\
& &SE &0.1527 &0.1315 &0.0926 &0.0771 &0.3025 &0.2563 &0.1758 &0.1477 \\
& &RE & 1.6698	&1.3274	&1.6879	&1.3448	&1.5078	&1.2412	&1.5147	&1.2368\\
&\multirow[t]{2}{*}{CAL} &Mean &0.7294 &0.7091 &0.7058 &0.7007 &0.7252 &0.7074 &0.7061 &0.6996 \\
& &SD &0.1732 &0.1412 &0.0963 &0.0786 &0.3155 &0.2557 &0.1793 &0.1510 \\
& &RE & 1.6527&	1.3966	&1.6603	&1.3482	&1.4946	&1.2311	&1.5029	&1.2678 \\
&\multirow[t]{2}{*}{BSc} &Mean &0.7245 &0.7074 &0.7065 &0.7026 &0.7357 &0.7070 &0.7051 &0.7008 \\
& &SD &0.1742 &0.1344 &0.0977 &0.0783 &0.3181 &0.2577 &0.1805 &0.1473 \\
& &RE & 1.6622	&1.3294&1.6845	&1.3431	&1.5069	&1.2407&1.5130	&1.2368\\
\bottomrule
\end{tabular*}
\end{table}

\begin{sidewaystable}[!htp] \centering
\caption{Stratified sampling design implemented for NWTS and corresponding cohort and sample sizes for each stratum.}
\label{tab: StratiDes}
\scriptsize
\begin{tabular*}{\textwidth}{@{\extracolsep\fill}c*{9}{p{1.2cm}}@{}}\toprule 
&\multirow{3}{*}{Total} &\multicolumn{4}{c}{Favorable Histology} &\multicolumn{4}{c}{Unfavorable Histology} \\\cmidrule(lr){3-6}   \cmidrule(lr){7-10} 
& &\multicolumn{2}{c}{Stage I, II} &\multicolumn{2}{c}{Stage II, IV} &\multicolumn{2}{c}{Stage I, II} &\multicolumn{2}{c}{Stage II, IV} \\\cmidrule(lr){3-4}  \cmidrule(lr){5-6} \cmidrule(lr){7-8} \cmidrule(lr){9-10}
& &Age $<$ 1 Year &Age $\geq$ 1 Year &Age $<$ 1 Year &Age $\geq$ 1 Year &Age $<$ 1 Year &Age $\geq$ 1 Year &Age $<$ 1 Year &Age $\geq$ 1 Year \\\midrule
&\multicolumn{9}{c}{Main Study Cohort or Phase 1 Sample ($N = 3915$)} \\
Cases &669 &51 &238 &7 &211 &13 &43 &28 &78 \\
Controls &3246 &397 &1675 &28 &926 &11 &108 &2 &99 \\
\% Relapse &17.09\% &11.38\% &12.44\% &20.00\% &18.56\% &54.17\% &28.48\% &93.33\% &44.07\% \\
&\multicolumn{9}{c}{Phase 2 Sample ($n = 1317$)} \\
Cases &669 &51 &238 &7 &211 &13 &43 &28 &78 \\
Controls &648 &120 &160 &28 &120 &11 &108 &2 &99 \\
\bottomrule
\end{tabular*}
\end{sidewaystable}

\begin{sidewaystable}[!htp]\centering
\caption{Results for $2000$ simulated phase 2 samples from NWTS. SRS, BS, CAL, and BSc denote simple random sampling, balanced sampling, calibration, and re-calibration after balanced sampling, respectively. UH denotes the unfavorable central path histology. Age0 and Age 1 denote piecewise linear terms for age at diagnosis (years) before and after one year, respectively. Stage denotes a binary indicator of stage I-II disease, and Diameter denotes the diameter of the tumor. $\tilde{\beta}_N$ and $\text{SE}_1$ denote the Cox model coefficients and standard errors for the full cohort, respectively. Mean denotes the mean of the estimated regression coefficients, SD denotes the standard deviation of the estimated regression coefficients, SE denotes the mean of the estimated standard errors, and RE denotes the relative efficiency.} \label{tab: nwts_phase2_stratification}
\footnotesize
\begin{tabular}{ccccccccccccccccc}\toprule
\multirow{3}{*}{Model Term} &\multicolumn{2}{c}{Phase 1 Estimates} &\multicolumn{13}{c}{Summary Statistics for Phase 2 Estimates} \\\cmidrule(lr){2-3} \cmidrule(lr){4-16}
&\multirow{2}{*}{$\tilde{\beta}_N$} &\multirow{2}{*}{$\text{SE}_1$} &\multicolumn{3}{c}{SRS} &\multicolumn{4}{c}{BS} &\multicolumn{3}{c}{CAL} &\multicolumn{3}{c}{BSc} \\\cmidrule(lr){4-6} \cmidrule(lr){7-10} \cmidrule(lr){11-13} \cmidrule(lr){14-16}
& & &Mean &SD &RE &Mean &SD &SE &RE &Mean &SD &RE &Mean &SD &RE \\\midrule
UH &4.0418 &0.1708 &3.8560 &0.1657 &0.9701 &4.0556 &0.1298 &0.1164 &0.7600 &3.9907 &0.1238 &0.7248 &4.0564 &0.1285 &0.7523 \\
Age0 &-0.6608 &0.1060 &-0.6668 &0.1415 &1.3349 &-0.6629 &0.0435 &0.0440 &0.4104 &-0.6585 &0.0377 &0.3557 &-0.6600 &0.0353 &0.3330 \\
Age1 &0.1041 &0.0003 &0.1011 &0.0195 &65.0000 &0.1044 &0.0075 &0.0073 &25.0000 &0.1133 &0.0072 &24.0000 &0.1039 &0.0059 &19.6667 \\
Stage &-1.3463 &0.0594 &-1.3351 &0.2201 &3.7054 &-1.3395 &0.0830 &0.0977 &1.3973 &-1.3137 &0.0666 &1.1212 &-1.3366 &0.0700 &1.1785 \\
Diameter &-0.0063 &0.0002 &-0.0183 &0.0139 &69.5000 &-0.0054 &0.0049 &0.0043 &24.5000 &-0.0095 &0.0033 &16.5000 &-0.0054 &0.0038 &19.0000 \\
Stage $\times$ Diameter &0.0756 &0.0004 &0.0725 &0.0195 &48.7500 &0.0752 &0.0070 &0.0081 &17.5000 &0.0742 &0.0060 &15.0000 &0.0749 &0.0058 &14.5000 \\
UH $\times$ Age0 &-2.6354 &0.2153 &-2.2786 &0.2428 &1.1277 &-2.6669 &0.2403 &0.3189 &1.1161 &-2.4212 &0.2117 &0.9833 &-2.6690 &0.2392 &1.1110 \\
UH $\times$ Age1 &-0.0577 &0.0011 &-0.0886 &0.0431 &39.1818 &-0.0487 &0.0466 &0.0581 &42.3636 &-0.0957 &0.0439 &39.9091 &-0.0483 &0.0465 &42.2727 \\
\bottomrule
\end{tabular}
\end{sidewaystable}


\begin{figure}[ht!]
    \centering
    \includegraphics[width=\textwidth]{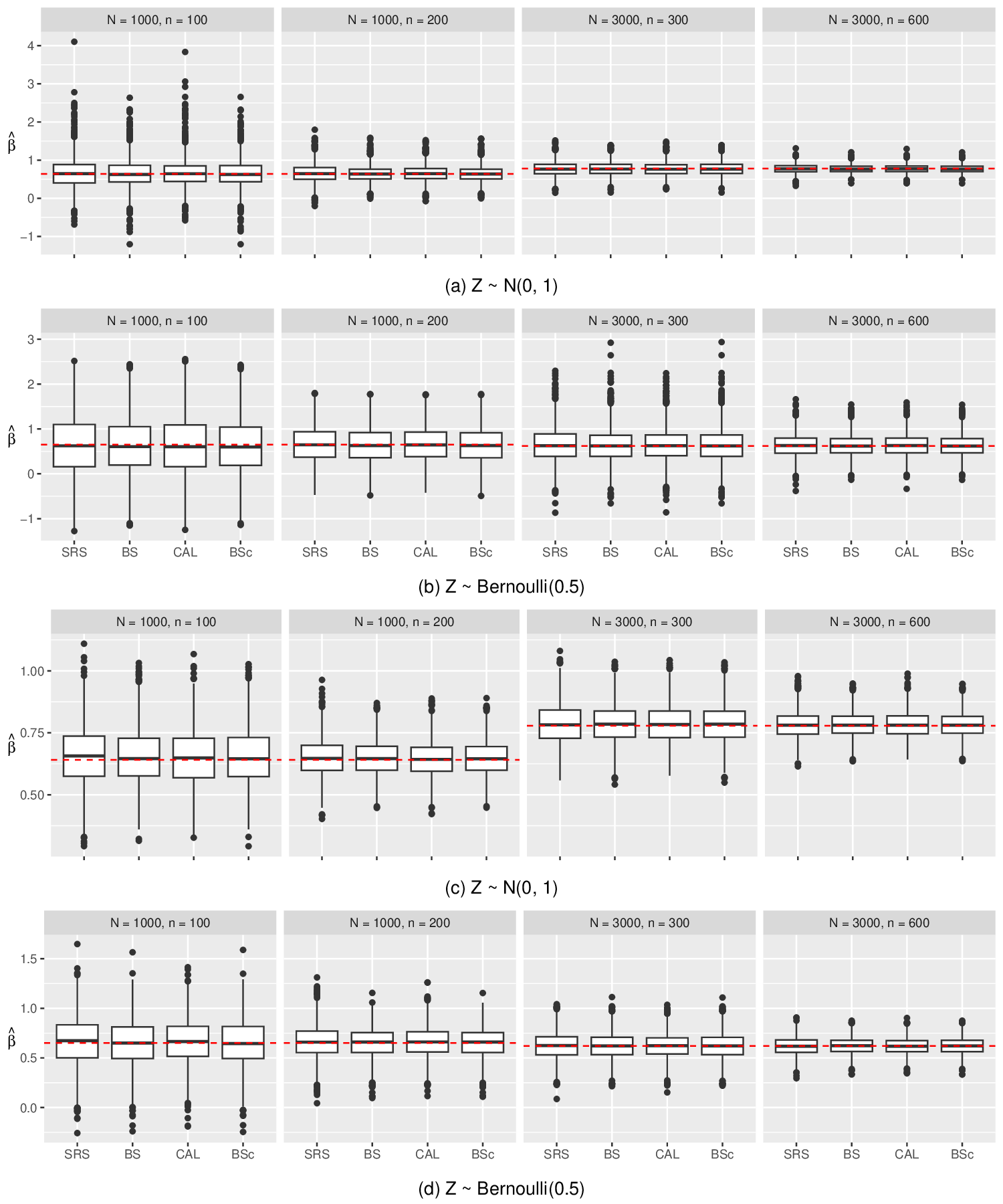}
    \caption{Boxplots of estimated regression coefficients for a fixed cohort with the censoring proportion of $90\%$ and correlation between covariates being $\rho=0.5$ for Simulation Setup I. The red-dotted horizontal line represents the estimated regression coefficient based on the full cohort data. (a) refers to the results from subcohort samplings for continuous covariates, (b) refers to the results from subcohort samplings for binary covariates, (c) refers to the results from case-cohort samplings for continuous covariates, and (d) refers to the results from case-cohort samplings for binary covariates. $N$ denotes the cohort size, $n$ denotes the subcohort size, and $\hat{\beta}$ denotes the estimated regression coefficients. SRS, BS, CAL, and BSc denote simple random sampling, balanced sampling, calibration, and re-calibration after balanced sampling, respectively.}
    \label{fig: fix_0.5}
\end{figure}

\begin{figure}[ht!]
    \centering
    \includegraphics[width=\textwidth]{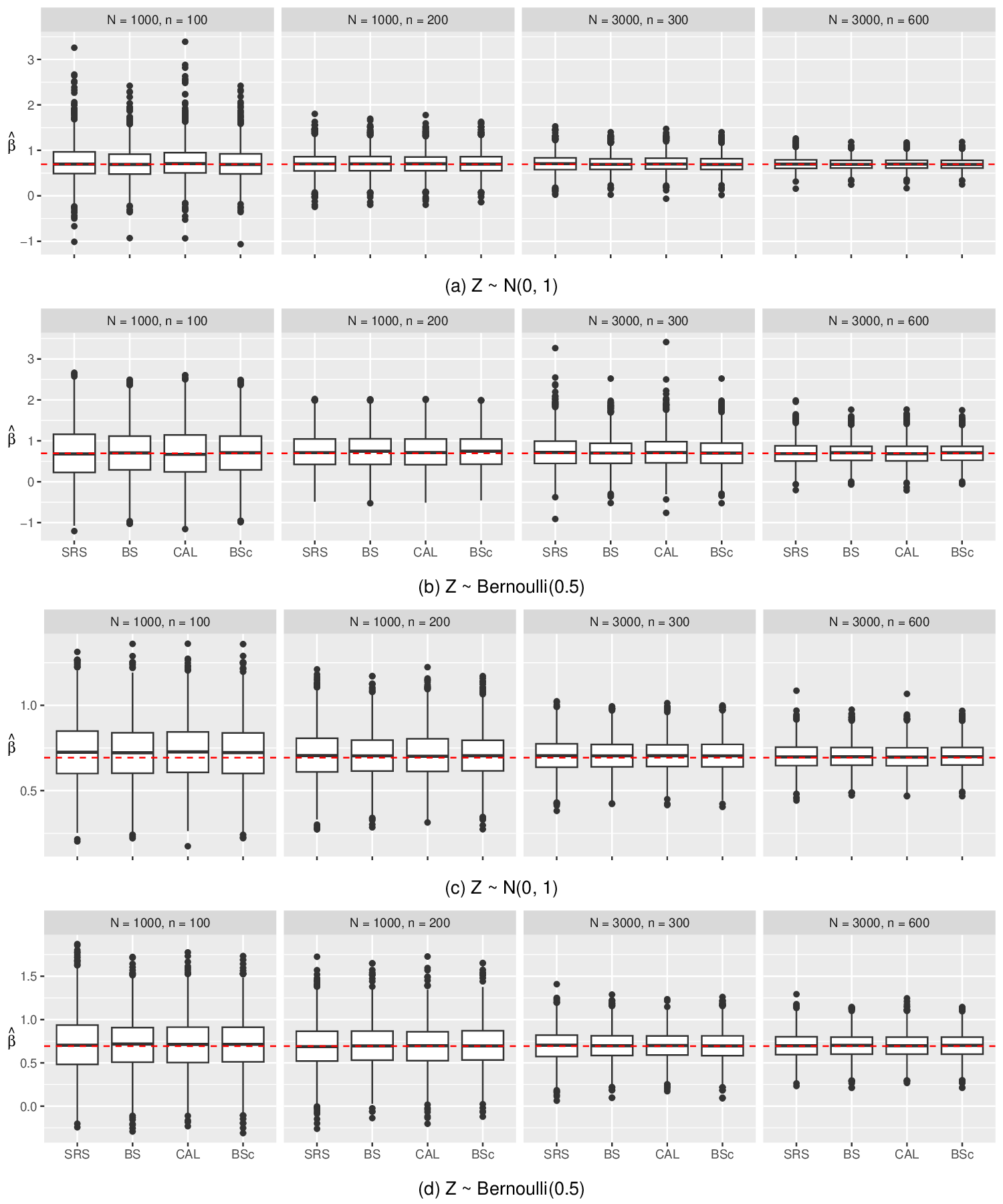}
    \caption{Boxplots of estimated regression coefficients for random cohorts with the censoring proportion of $90\%$ and correlation between covariates being $\rho=0.5$ for Simulation Setup II. The red-dotted horizontal line represents the estimated regression coefficient based on the full cohort, $\log{2}$. (a) refers to the results from subcohort samplings for continuous covariates, (b) refers to the results from subcohort samplings for binary covariates, (c) refers to the results from case-cohort samplings for continuous covariates, and (d) refers to the results from case-cohort samplings for binary covariates. $N$ denotes the cohort size, $n$ denotes the subcohort size, and $\hat{\beta}$ denotes the estimated regression coefficients. SRS, BS, CAL, and BSc denote simple random sampling, balanced sampling, calibration, and re-calibration after balanced sampling, respectively.}
    \label{fig: random_0.5}
\end{figure}


\end{document}